\documentclass[prl, twocolumn,superscriptaddress,amsmath,amssymb,showpacs,floatfix,preprintnumbers]{revtex4-2}

\usepackage{amsmath}
\usepackage{graphicx}
\usepackage{dcolumn}
\usepackage{xcolor}
\usepackage{physics}
\usepackage{xr}
\usepackage{siunitx}
\usepackage{tikz}
\usepackage{tabularx}
\usepackage{makecell}
\usepackage{multirow}
\DeclareGraphicsExtensions{.png .jpg .pdf}

\usepackage[normalem]{ulem}

\usepackage{pifont}
\newcommand{\cmark}{\ding{51}}%
\newcommand{\xmark}{\ding{55}}%

\makeatletter
\newcommand*{\addFileDependency}[1]{
  \typeout{(#1)}
  \@addtofilelist{#1}
  \IfFileExists{#1}{}{\typeout{No file #1.}}
}
\makeatother
 
\newcommand*{\myexternaldocument}[1]{%
    \externaldocument{#1}%
    \addFileDependency{#1.tex}%
    \addFileDependency{#1.aux}%
}

\myexternaldocument{si}

\begin{document}

\title{Linear Magnetoelectric Electro-Optic Effect}

\author{D. J. P. de Sousa}
\affiliation{Department of Electrical and Computer Engineering, University of Minnesota, Minneapolis, Minnesota 55455, USA}
\author{C. O. Ascencio}
\affiliation{School of Physics and Astronomy, University of Minnesota, Minneapolis, Minnesota 55455, USA}
\author{Tony Low}\email{tlow@umn.edu}
\affiliation{Department of Electrical and Computer Engineering, University of Minnesota, Minneapolis, Minnesota 55455, USA}
\affiliation{School of Physics and Astronomy, University of Minnesota, Minneapolis, Minnesota 55455, USA}

\begin{abstract}
In this work, we derive a generalized constitutive relation describing the current response to external electromagnetic fields in electrically biased quantum materials. While our semiclassical Boltzmann approach reveals the existence of electro-optic effects induced by the Berry curvature dipole of Bloch electrons, we also find a wealth of alternative electro-optic effects originating from the interplay between Berry curvature and magnetic moment. In particular, our symmetry analysis indicates the existence of a \textit{magnetoelectric electro-optic effect}, derived from the simultaneous presence of Berry curvature and magnetic moment, that requires either time-reversal or inversion symmetry breaking. The revealed conductivity coefficients are explicitly written and we derive the tensor shape describing such alternative electromagnetic responses for chiral materials pertaining to space groups 152 and 198.

\end{abstract}

\maketitle

\section{I. Introduction}
The Berry curvature of Bloch electrons gives rise to a diverse array of alternative electromagnetic responses in quantum materials~\cite{RevModPhys.82.1959, Ma2023, PhysRevLett.130.076901, PhysRevB.99.155404, PhysRevLett.115.216806, deJuan2017, PhysRevB.97.035158, PhysRevB.97.205206, PhysRevB.92.235205, PhysRevB.105.125201, https://doi.org/10.48550/arxiv.2312.15142, https://doi.org/10.48550/arxiv.2401.13764, PhysRevB.107.125151, Dai2023}. Recently, significant efforts have been devoted to elucidating the influence of the Berry curvature dipole (BCD) on the electromagnetic properties of non-centrosymmetric systems, resulting in a profound understanding on the nature of non-linear bulk photogalvanic effects and linear electro-optic effects~\cite{Ma2023, PhysRevLett.115.216806, PhysRevLett.105.026805, PhysRevB.102.085202, PhysRevLett.130.076901, https://doi.org/10.48550/arxiv.2401.13764, PhysRevB.92.235447}. For instance, the presence of a weak static electric field $\textbf{E}_0$ was recently shown to modify the optical conductivity of non-centrosymmetric systems as
\begin{eqnarray}
\sigma_{\omega E}^{\alpha\beta} \rightarrow \sigma_{\omega E}^{\alpha\beta} + \sigma_{\omega E}^{\alpha\beta\gamma}E_0^{\gamma},
    \label{eq00000}
\end{eqnarray}
where $\sigma_{\omega E}^{\alpha\beta\gamma}$ originates from the BCD on the Fermi surface, giving rise to gyrotropic effects~\cite{PhysRevLett.130.076901}. Beyond practical importance, these advances highlight the potential hidden in the wave function of Bloch electrons for emerging alternative electromagnetic responses in quantum materials, a timely topic with important ramifications to the field of optoelectronics~\cite{Ma2023}.

Parallel to this research activity, intriguing electromagnetic responses originating from the magnetic moment texture of Bloch electrons, arising from their intrinsic spin and orbital angular momentum, have been investigated~\cite{PhysRevLett.95.137205, PhysRevB.85.014435, PhysRevB.94.245121, Malashevich2010, Vanderbilt2018, PhysRevB.96.035120, PhysRevLett.116.077201, PhysRevB.101.174419, PhysRev.171.1065, PhysRevB.82.245118, PhysRevB.98.155145}. A prominent example is the prediction of a gyrotropic magnetic effect~\cite{PhysRevLett.116.077201, PhysRevB.92.235205}, where a charge current, $\textbf{J}_{\omega}$, is generated in response to an AC magnetic field, $\textbf{B}_{\omega}$, resulting in the relation
\begin{eqnarray}
J^{\alpha}_{\omega} = \sigma^{\alpha\beta}_{\omega B}B^{\beta}_{\omega},
\label{eq_JsB}
\end{eqnarray}
with the frequency-dependent coefficient, $\sigma^{\alpha\beta}_{\omega B}$, capturing the magnetic moment distribution on the Fermi surface. Here, $\sigma^{\alpha\beta}_{\omega B}$ also describes a dynamical magnetoelectric coupling~\cite{PhysRevLett.116.077201}, expressed as
\begin{subequations}
\begin{eqnarray}
  P_{\omega}^{\alpha} = \displaystyle \frac{i}{\omega} \sigma_{\omega B}^{\alpha\beta} B_{\omega}^{\beta} ,
\end{eqnarray}

\begin{eqnarray}
  M_{\omega}^{\beta} = \displaystyle -\frac{i}{\omega} \sigma_{\omega B}^{\alpha\beta}  E_{\omega}^{\alpha} , 
\end{eqnarray}
\end{subequations}
for the polarization, $\textbf{P}_{\omega}$, and magnetic moment, $\textbf{M}_{\omega}$, induced by an oscillating electromagnetic field.

While these developments examined the Berry curvature and magnetic moment of Bloch electrons in an independent manner, it is not clear whether their concomitant presence would produce unexpected optoelectronic phenomena. An intriguing possibility is the emergence of new optical phenomenon due to the simultaneous presence of these quantities, thus offering a rich playground for theoretical and experimental exploration. 

\begin{figure}[t]
\centerline{\includegraphics[width =\linewidth]{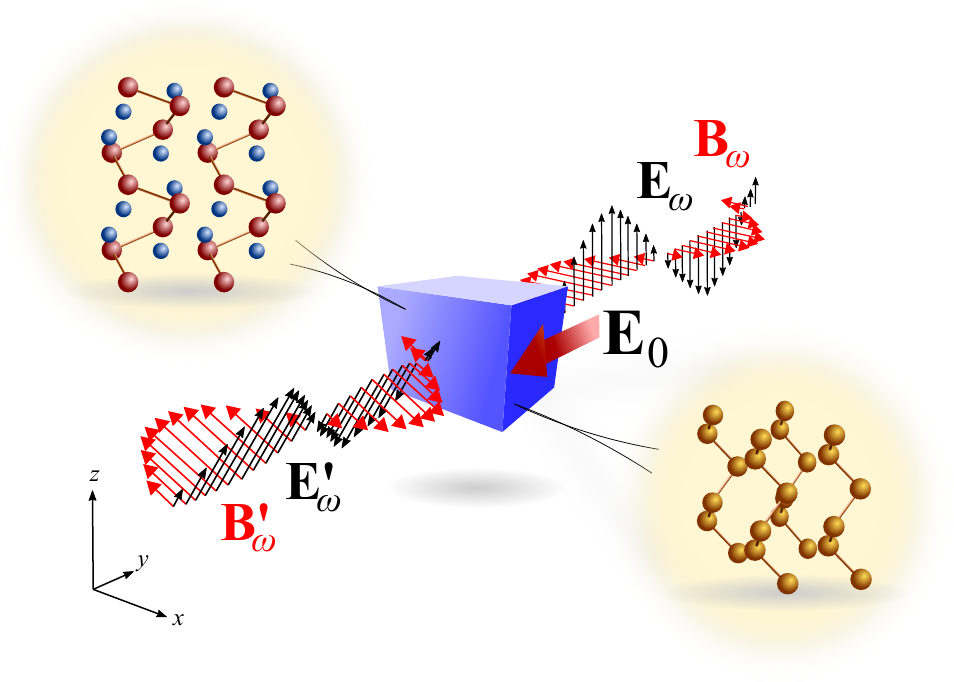}}
\caption{Electro-optic effects in quantum materials, e.g., a rotation of the plane of polarization of light can be induced or amplified by means of an applied static electric field $\textbf{E}_0$. The wave function of Bloch electrons in quantum materials, such as non-centrosymmetric chiral crystals, support quantities that enable alternative electro-optic effects.}
\label{Fig1}
\end{figure}

In this paper, we investigate the electromagnetic responses in metals emerging from coexisting Berry curvature and magnetic moment texture of Bloch electrons, focusing on electro-optic effects~\cite{PhysRevLett.130.076901, 1987, Weber2018, PhysRevLett.125.017401}; The presence of an static bias $\textbf{E}_0$ modifies the electromagnetic properties of quantum materials, enabling the manipulation of scattered electromagnetic waves, as depicted in Fig.~\ref{Fig1}. Here, we employ the semi-classical Boltzmann formalism to derive the full current response of Bloch electrons subjected to AC electromagnetic fields, when $\textbf{E}_0$ is present. Our results reveal the existence of a unique electromagnetic signature originating from coexisting magnetic moment and Berry curvature texture in quantum materials with either broken inversion or time-reversal symmetry. Such signature can be understood as the first order correction in $\textbf{E}_0$ field to the dynamical magnetoelectric coupling coefficient of Eq.~(\ref{eq_JsB}), such that
\begin{eqnarray}
\sigma_{\omega B}^{\alpha\beta} \rightarrow \sigma_{\omega B}^{\alpha\beta} + \sigma_{\omega B}^{\alpha\beta\gamma}E_0^{\gamma},
    \label{eq0000}
\end{eqnarray}
holds true in the weak bias limit. Further, the existence of this effect in non-centrosymmetric systems recovers the symmetry of the current response to electromagnetic fields, ($\textbf{E}_{\omega}$, $\textbf{B}_{\omega}$), in the presence of bias;
\begin{eqnarray}
    J^{\alpha}(\omega) &= (\sigma_{\omega E}^{\alpha\beta} + \sigma_{\omega E}^{\alpha\beta\gamma}E_0^{\gamma}) E^{\beta}_{\omega}  + (\sigma_{\omega B}^{\alpha\beta} + \sigma_{\omega B}^{\alpha\beta\gamma}E_0^{\gamma}) B^{\beta}_{\omega}, \nonumber \\
\end{eqnarray}
to first order in $|\textbf{E}_0|$. Thus, $\sigma_{\omega B}^{\alpha\beta\gamma}$ is the natural magnetoelectric analogue of the BCD contribution $\sigma_{\omega E}^{\alpha\beta\gamma}$ under static bias. For the above reason, we refer to this particular electromagnetic signature as a \textit{magnetoelectric electro-optic effect} in this work. 

We further analyse the role of symmetry in determining the shape of the associated electro-optic tensors for materials classified within space groups (SG)-152 and 198, such as chiral Te and Se\cite{Qiu2022, PhysRevLett.124.136404, PhysRevLett.125.216402, PhysRevB.97.035158}, as well as the topological chiral CoSi, RhSi~\cite{Chang2018, Bradlyn2016, PhysRevLett.130.066402, Rao2019, Schrter2019, PhysRevLett.119.206401, PhysRevB.108.L201404, PshenaySeverin2018, PhysRevLett.119.206402, PhysRevB.98.155145}. We show that such systems must display magnetoelectric electro-optic effects, described through a fully diagonal tensor, offering clear guidance to the experimental probing of its optical signatures.

The paper is organized as follows. In Sec. II we account for the semiclassical Boltzmann formalism adopted in this work. In Sec. III we describe, in a point-by-point manner, the various contributions to the non-equilibrium distribution function consistent with the semiclassical approach. The transport coefficients describing the full linear response are explicitly given in Secs. IV and V, where the former summarizes the DC-field independent generalized constitutive relation, while the later focuses on the description of electro-optic effects. We apply the developed theory to the case of SG-152 and SG-198 materials in Sec. VI, where we present the symmetry-imposed shaped of the various tensors describing electro-optic effects. We conclude in Sec. VII with a brief summary.

\section{II. Formalism}\label{sectionII}

In this section, we describe the basic formalism adopted in this work. Here, we establish our conventions and approximations in regards to the linear response framework and semiclassical Boltzmann formalism.  

We assume a material system biased with an electric field $\textbf{E}_0$. For weak AC field variations around the DC field, we write 
\begin{eqnarray}
    \textbf{E} = \textbf{E}_0 + \textbf{E}_{\omega}e^{-i\omega t} + \textbf{E}^*_{\omega}e^{i\omega t},
    \label{eq1}
\end{eqnarray}
\begin{eqnarray}
    \textbf{B} = \textbf{B}_{\omega}e^{-i\omega t} + \textbf{B}^*_{\omega}e^{i\omega t},
    \label{eq2}
\end{eqnarray}
and the electron response is linear in $\textbf{E}_{\omega}$ and $\textbf{B}_{\omega}$. In this work, $\textbf{E}_{\omega},\textbf{B}_{\omega} \ll \textbf{E}_0$, such that $\textbf{E}^2_{\omega} (\textbf{B}^2_{\omega})$ contribution is negligible and $\textbf{E}^2_0 $ is beyond linear response, but responses involving products of $\textbf{E}_{0}$ and $\textbf{B}_{\omega}$ and, likewise, $\textbf{E}_{0}$ and $ \textbf{E}_{\omega}$, are still sizable and non-negligible within the linear response theory framework. 

We employ the Boltzmann formalism to study the system's electromagnetic response. The semiclassical equations of motion in the presence of an electromagnetic field are~\cite{PhysRevB.88.104412}
\begin{eqnarray}
    \textbf{\.x}_{n\textbf{k}} = \textbf{v}_{n\textbf{k}} - \textbf{\.k}\times\boldsymbol{\Omega}_{n\textbf{k}},
    \label{eq001}
\end{eqnarray}
\begin{eqnarray}
    \hbar\textbf{\.k} = -e\textbf{E} - e \textbf{\.x}_{n\textbf{k}} \times \textbf{B},
    \label{eq002}
\end{eqnarray}
where $\boldsymbol{\Omega}_{n\textbf{k}}$ is the Berry curvature associated with the Bloch state, $|u_{n\textbf{k}}\rangle$ with energy $\epsilon_{n\textbf{k}}$, given by
\begin{eqnarray}
\boldsymbol{\Omega}_{n\textbf{k}} = -\operatorname{Im}\langle \nabla_{\textbf{k}} u_{n\textbf{k}}|\times |\nabla_{\textbf{k}}u_{n\textbf{k}}\rangle,
    \label{eq03}
\end{eqnarray}
and $\textbf{v}_{n\textbf{k}} = (1/\hbar)\nabla_{\textbf{k}}\epsilon_{n\textbf{k}}$ is the band velocity. The finite magnetic moment of Bloch electrons is captured through a Zeeman-like energy shift $\epsilon_{n\textbf{k}} \rightarrow \epsilon_{n\textbf{k}} - \textbf{m}_{n\textbf{k}}\cdot\textbf{B}$ and the corresponding band velocity correction $\hbar\textbf{v}_{n\textbf{k}} \rightarrow \hbar\textbf{v}_{n\textbf{k}} - \nabla_{\textbf{k}}(\textbf{m}_{n\textbf{k}}\cdot \textbf{B})$. The total magnetic moment of Bloch electrons, $\textbf{m}_{n\textbf{k}} = \textbf{L}_{n\textbf{k}} + \textbf{S}_{n\textbf{k}}$, is generally comprised of a spin contribution, $\textbf{S}_{n\textbf{k}} = -(eg_s\hbar/4m) \langle u_{n\textbf{k}}| \boldsymbol{\sigma} | u_{n\textbf{k}}\rangle$ where $g_s$ is the spin g-factor and $m$ the electron mass, and an orbital contribution~\cite{PhysRevLett.95.137205, PhysRevB.85.014435, PhysRevB.94.245121, Malashevich2010, Vanderbilt2018}  
\begin{eqnarray}
    \textbf{L}_{n\textbf{k}} = \displaystyle \frac{e}{2\hbar} \operatorname{Im}\langle \grad_{\textbf{k}} u_{n\textbf{k}} | \times (H_{\textbf{k}} - \epsilon_{n\textbf{k}})| \grad_{\textbf{k}} u_{n\textbf{k}}\rangle.
    \label{eq}
\end{eqnarray}

These two contributions to the total magnetic moment of Bloch electrons are typically described through a momentum space texture at a given energy, as specified by the vector fields $\textbf{S}_{n\textbf{k}}$ and $\textbf{L}_{n\textbf{k}}$~\cite{Yang2023, Go2017}. In upcoming sections, we discuss how different types of optical responses might depend on a particular magnetic moment texture of Bloch electrons.

We proceed by decoupling Eqs.~(\ref{eq001}) and (\ref{eq002}) and writing down the spatially homogeneous Boltzmann equation of the problem in the form [See Appendix II]
\begin{widetext}
\begin{eqnarray}
    \frac{\partial g_{n\textbf{k}}}{\partial t} + \left[ -e\textbf{E}\cdot \textbf{v}_{n\textbf{k}} - \frac{e^2}{\hbar} [\textbf{B}\times (\boldsymbol{\Omega}_{n\textbf{k}} \times \textbf{E})]\cdot \textbf{v}_{n\textbf{k}} + \frac{e}{\hbar}\textbf{B}\cdot \textbf{K}_{n\textbf{k}}\cdot \textbf{E}\right]\frac{\partial f^0_{n\textbf{k}}}{\partial \epsilon_{n\textbf{k}}} = -\frac{\delta g_{n\textbf{k}}}{\tau},
    \label{BoltzmnEq}
\end{eqnarray}    
\end{widetext}
within the relaxation time approximation, characterized by $\tau$. We have defined $\textbf{K}_{n\textbf{k}}$ to represent the Jacobian of the vector field $\textbf{m}_{n\textbf{k}}$, with components $K^{\alpha\beta}_{n\textbf{k}} = \nabla^{\beta}_{\textbf{k}}m^{\alpha}_{n\textbf{k}}$. Equation~(\ref{BoltzmnEq}) is the most general homogeneous Boltzmann equation consistent with our linear response framework. It accounts for the coupling between $\textbf{E}$ and $\textbf{B}$ fields brought about by the Berry curvature and magnetic moment of Bloch electrons. In particular, the Berry curvature contribution differs from the conventional $(\textbf{E}\cdot\textbf{B})\boldsymbol{\Omega}_{n\textbf{k}}$ term, associated with chiral anomaly in Weyl semimetals~\cite{PhysRevB.88.104412}. This is due to a first order correction to the phase-space volume $\propto (\boldsymbol{\Omega}_{n\textbf{k}}\cdot\textbf{B})$ that is still consistent with our approximations. In fact, $\textbf{B}\times (\boldsymbol{\Omega}_{n\textbf{k}} \times \textbf{E}) \approx (\textbf{E}\cdot\textbf{B})\boldsymbol{\Omega}_{n\textbf{k}}$ when $(\boldsymbol{\Omega}_{n\textbf{k}}\cdot\textbf{B}) \approx 0$, recovering the typical $\textbf{E}$ and $\textbf{B}$ coupling through the Berry curvature. A detailed derivation is presented in Appendix~\ref{AppII}. 

In what follows, we solve the Boltzmann equation, Eq.~(\ref{BoltzmnEq}), by taking into account Eqs.~(\ref{eq1}) and (\ref{eq2}). The final form of the non-equilibrium distribution function, $g_{n\textbf{k}}$, is at most linear in $\textbf{E}_{\omega}$ and $\textbf{B}_{\omega}$, with contributions that might display static field dependencies, e.g., arising from cross $\textbf{E}\cdot\textbf{B}$, such as $\textbf{E}_0\cdot\textbf{B}_{\omega}$ or $\textbf{E}_{0}\cdot\textbf{E}_{\omega}$.

\section{III. Non-equilibrium distribution function}\label{sectionIII}

In this section, we show explicitly the impact of DC and AC fields to the equilibrium distribution function of Bloch electrons. The solution to the spatially homogeneous Boltzmann equation renders the following non-equilibrium distribution function
\begin{eqnarray}
    g_{n\textbf{k}} = g^0_{n\textbf{k}} + \delta g^{\omega}_{n\textbf{k}}e^{-i\omega t} + \delta g^{\omega *}_{n\textbf{k}}e^{i\omega t},
    \label{eq004}
\end{eqnarray}
with static and dynamical contributions $g^0_{n\textbf{k}}$ and $\delta g^{\omega}_{n\textbf{k}}$, respectively, which we discuss in more detail in the following. 

We begin by addressing the the static contribution, $\ g^{0}_{n\textbf{k}}$. We find that it assumes the simple form $g^0_{n\textbf{k}} = f^0_{n\textbf{k}} + \delta g^{E_0}_{n\textbf{k}}$, where $f^0_{n\textbf{k}}$ is the equilibrium Fermi-Dirac distribution of Bloch electrons, and the non-equilibrium portion reads 
\begin{eqnarray}
    \delta g^{E_0}_{n\textbf{k}} = \frac{\partial f^0_{n\textbf{k}}}{\partial \epsilon_{n\textbf{k}}} e \tau \textbf{v}_{n\textbf{k}} \cdot \textbf{E}_0,
    \label{eq005}
\end{eqnarray}
corresponding to the usual DC Drude response to $\textbf{E}_0$. Next, we address the dynamical part of the non-equilibrium distribution function.

We find that the impact of the dynamical fields can most generally be separated into four distinct contributions, $\delta g^{\omega}_{n\textbf{k}} = \delta g^{E_{\omega}}_{n\textbf{k}} + \delta g^{B_{\omega}}_{n\textbf{k}} + \delta g^{\boldsymbol{\Omega}-E_0B_{\omega}}_{n\textbf{k}} + \delta g^{\textbf{m}-E_0B_{\omega}}_{n\textbf{k}}$.  

The first AC contribution, 
\begin{eqnarray}
    \delta g^{E_{\omega}}_{n\textbf{k}} = \frac{\partial f^0_{n\textbf{k}}}{\partial \epsilon_{n\textbf{k}}} \frac{e \tau}{1 - i\omega\tau} \textbf{v}_{n\textbf{k}} \cdot \textbf{E}_{\omega},
    \label{eq6}
\end{eqnarray}
corresponds to the typical AC Drude response, and describes an out-of-equilibrium correction to the distribution of Bloch electrons brought about by the oscillating electric field, $\textbf{E}_{\omega}$. 

The second AC contribution, 
\begin{eqnarray}
    \delta g^{B_{\omega}}_{n\textbf{k}} = \frac{\partial f^0_{n\textbf{k}}}{\partial \epsilon_{n\textbf{k}}} \frac{i\omega\tau}{i\omega\tau - 1} \textbf{m}_{n\textbf{k}} \cdot \textbf{B}_{\omega},
    \label{eq77}
\end{eqnarray}
describes how the AC magnetic field perturbs the distribution of states on the Fermi surface, through its coupling to the magnetic moment of Bloch electrons. Such contribution has been theoretically predicted to give rise to charge currents responses collinear with $\textbf{B}_{\omega}$~\cite{PhysRevLett.116.077201}.

Next, we have
\begin{eqnarray}
    \delta g^{\boldsymbol{\Omega}-E_0B_{\omega}}_{n\textbf{k}} = \frac{\partial f^0_{n\textbf{k}}}{\partial \epsilon_{n\textbf{k}}} \frac{e^2 \tau/\hbar}{1 - i\omega\tau} [\textbf{B}_{\omega}\times (\boldsymbol{\Omega}_{n\textbf{k}} \times \textbf{E}_{0})]\cdot \textbf{v}_{n\textbf{k}}, \nonumber \\
    \label{eq88}
\end{eqnarray}
Such contribution, arising from the $\textbf{B}\times (\boldsymbol{\Omega}_{n\textbf{k}} \times \textbf{E})$ term appearing in the Boltzmann equation, is not typically taken into account in most cases. We find, however, that we cannot neglect these contributions in order to maintain consistency with the initial assumptions concerning the linear response framework.

Finally, the fourth AC contribution to the non-equilibrium distribution function reads
\begin{eqnarray}
    \delta g^{\textbf{m}-E_0B_{\omega}}_{n\textbf{k}} = -\frac{\partial f^0_{n\textbf{k}}}{\partial \epsilon_{n\textbf{k}}} \frac{e \tau/\hbar}{1 - i\omega\tau} [\textbf{B}_{\omega}\cdot \textbf{K}_{n\textbf{k}}\cdot \textbf{E}_0],
    \label{eq008}
\end{eqnarray}
originating from the magnetic moment of Bloch electrons on the Fermi surface, expressed through $\textbf{K}_{n\textbf{k}}$. We refer the reader to Appendix~\ref{AppIII} for the full derivation of the various contributions described above. Once established the full form of the non-equilibrium distribution function of Bloch electrons, we proceed by deriving the full electromagnetic response.

\section{IV. Generalized constitutive relation}\label{sectionIV}

In this section, we write down a generalized constitutive relation describing the current response of Bloch electrons to static and dynamical electromagnetic fields. The charge current,~\cite{PhysRevB.88.104412, PhysRevB.89.195137, PhysRevB.107.014302}
\begin{eqnarray}
   \textbf{J} = -e\displaystyle\sum_{n\textbf{k}}  g_{n\textbf{k}} \left[\textbf{v}_{n\textbf{k}} + \left(\frac{e}{\hbar}\right)\textbf{E}\times\boldsymbol{\Omega}_{n\textbf{k}} + \left(\frac{e}{\hbar }\right)(\boldsymbol{\Omega}_{n\textbf{k}} \cdot \textbf{v}_{n\textbf{k}})\textbf{B}\right], \nonumber \\
    \label{eq003}
\end{eqnarray}
is expressed in terms of the non-equilibrium distribution function, $g_{n\textbf{k}}$, derived in the previous section. Here, we emphasize that the band velocity should be interpreted as $\hbar\textbf{v}_{n\textbf{k}} \rightarrow \hbar\textbf{v}_{n\textbf{k}} - \nabla_{\textbf{k}}(\textbf{m}_{n\textbf{k}}\cdot \textbf{B})$. Combining Eq.~(\ref{eq003}) with Eqs.~(\ref{eq004})-(\ref{eq008}), and accounting for Eqs.~(\ref{eq1}) and (\ref{eq2}), we obtain a current response separable into static and dynamical contributions, i.e., $\textbf{J}(\omega) = \textbf{J}_0 + \textbf{J}_{\omega}e^{-i\omega t} + \textbf{J}^{*}_{\omega}e^{i\omega t} + \mathcal{O}(2\omega)$. 

For the sake of completeness, we write next the DC current responses, $\textbf{J}_0$: 
\begin{eqnarray}
J^{\alpha}_{0} = \sigma^{\alpha\beta}_{0E}E^{\beta}_{0} + \epsilon_{\gamma\beta\alpha}\sigma^{\gamma}_{0HE}E^{\beta}_{0} 
    \label{main_result}
\end{eqnarray}
with transport coefficients 
\begin{subequations}
    \begin{eqnarray}
       \sigma^{\alpha\beta}_{0E} = e^2 \tau\displaystyle \sum_{n\textbf{k}} \left(- \frac{\partial f^0_{n\textbf{k}}}{\partial \epsilon_{n\textbf{k}}}  \right)  v^{\alpha}_{n\textbf{k}} v^{\beta}_{n\textbf{k}}, 
       \label{eq18a}
    \end{eqnarray}

    \begin{eqnarray}
       \boldsymbol{\sigma}_{0HE} = \displaystyle - \frac{e^2}{\hbar} \sum_{n\textbf{k}} f^0_{n\textbf{k}} \boldsymbol{\Omega}_{n\textbf{k}}, 
       \label{eq18b}
    \end{eqnarray}
\end{subequations}
corresponding to the regular Drude and anomalous Hall responses. While the Drude response is allowable in time-reversal symmetric and/or inversion symmetric systems, the anomalous Hall response requires time-reversal symmetry breaking by virtue of constraints imposed over the Berry curvature.

We note that the magnetic moment texture of Bloch electrons does not contribute to the DC current response within our linear response framework. On the other hand, our findings reveal that the AC linear responses are sensitive to the magnetic moment of Bloch electrons and depends also on the static electric field $\textbf{E}_0$. Next, we summarize the AC responses in a generalized constitutive relation. First, we discuss the static electric field-free responses and their dependencies on the magnetic moment texture of Bloch electrons and, then, we address the contributions arising from the presence of $\textbf{E}_0$.

\subsection{A. The Generalized AC Response}

The constitutive relation for the AC response, in the absence of a static electric field, can be written compactly as
\begin{eqnarray}
J^{\alpha}_{\omega} = \sigma^{\alpha\beta}_{\omega E}E^{\beta}_{\omega} + \epsilon_{\gamma\beta\alpha}\sigma^{\gamma}_{\omega HE}E^{\beta}_{\omega} +  \sigma^{\alpha\beta}_{\omega B} B_{\omega} + \epsilon_{\gamma\beta\alpha}\sigma^{\gamma}_{\omega HB} B^{\beta}_{\omega},\nonumber \\
    \label{main_result}
\end{eqnarray}
which is our first main formal result. Equation~(\ref{main_result}) describes the full current response to AC electromagnetic fields, within the linear response framework, accounting for the Berry curvature and magnetic moment of Bloch electrons. As a consequence, the full symmetry between electric and magnetic field responses are recovered, i.e., with the presence of ``longitudinal" and transverse responses to both $\textbf{E}_{\omega}$ and $\textbf{B}_{\omega}$. The transport coefficients are
 \begin{subequations}
    \begin{eqnarray}
    \sigma^{\alpha\beta}_{\omega E} = \frac{e^2\tau}{1-i\omega\tau}\sum_{n\textbf{k}} \left(-\frac{\partial f^0_{n\textbf{k}}}{\partial \epsilon_{n\textbf{k}}}\right)  v^{\alpha}_{n\textbf{k}}v^{\beta}_{n\textbf{k}},
        \label{eq18a}
    \end{eqnarray}
        \begin{eqnarray}
    \boldsymbol{\sigma}_{\omega HE} = \displaystyle -\frac{e^2}{\hbar}\sum_{n\textbf{k}}f^0_{n\textbf{k}} \boldsymbol{\Omega}_{n\textbf{k}} ,
        \label{eq18b}
    \end{eqnarray}
        \begin{eqnarray}
    &\sigma_{\omega B}^{\alpha\beta} =  \displaystyle-\frac{e^2}{\hbar}\sum_{n\textbf{k}} 
f_{n\textbf{k}}^0(\textbf{v}_{n\textbf{k}}\cdot \boldsymbol{\Omega}_{n\textbf{k}})\delta_{\alpha\beta} \nonumber \\
&\displaystyle + e \sum_{n\textbf{k}} \left(-\frac{\partial f^0_{n\textbf{k}}}{\partial \epsilon_{n\textbf{k}}}\right) m_{n\textbf{k}}^{\alpha} v_{n\textbf{k}}^{\beta} \nonumber \\
&\displaystyle + e \frac{i\omega\tau}{i\omega\tau - 1}\sum_{n\textbf{k}} \left(-\frac{\partial f^0_{n\textbf{k}}}{\partial \epsilon_{n\textbf{k}}}\right) m_{n\textbf{k}}^{\beta} v_{n\textbf{k}}^{\alpha},\nonumber \\
        \label{eq18c}
    \end{eqnarray}
        \begin{eqnarray}
    \boldsymbol{\sigma}_{\omega HB} = \displaystyle \frac{e}{\hbar}\sum_{n\textbf{k}} f_{n\textbf{k}}^0 (\nabla_{\textbf{k}}\times \textbf{m}_{n\textbf{k}}),
        \label{eq18d}
    \end{eqnarray}
\end{subequations}   
where $\delta_{\alpha\beta}$ is the Kronecker delta. Besides the regular AC Drude and anomalous Hall responses, our formalism indicates the presence of ``magnetoelectric-" and ``transverse magnetoelectric-" like AC responses, described by the $\boldsymbol{\sigma}_{\omega B}$ tensor and the $\boldsymbol{\sigma}_{\omega HB}$ vector, respectively. Equation~(\ref{eq18c}) is the sum of three distinct contributions; The Fermi sea term, $\propto \sum_{n\textbf{k}} f_{n\textbf{k}}^0(\textbf{v}_{n\textbf{k}}\cdot \boldsymbol{\Omega}_{n\textbf{k}})$, originates from the coupling between oscillating magnetic field and the Berry curvature of Bloch electrons that naturally arises from decoupling the semiclassical equations of motion~\cite{PhysRevB.88.104412}. It produces a charge current longitudinal with the oscillating magnetic field, whenever the integrand can be made finite. The theory of topological Weyl semimetals elaborates that such responses can arise from an imbalance between the population of right- and left-handed Weyl fermions whenever electric and magnetic fields are simultaneously present and satisfies $\textbf{E}\cdot\textbf{B} \neq 0$, a contribution often referred to as the chiral magnetic effect~\cite{PhysRevB.88.104412, PhysRevD.78.074033, PhysRevLett.111.027201, PhysRevB.92.161110}. The next two contributions to the AC magnetoelectric response depend fundamentally on the magnetic moment of Bloch electrons on the Fermi surface. While the frequency-independent contribution originates from the Zeeman correction to the band velocity of Bloch electrons, the frequency-dependent response arises from the Zeeman correction to the Bloch state energy through the non-equilibrium distributions function given in Eq.~(\ref{eq77}). The latter contribution has been derived in Ref.~\cite{PhysRevLett.116.077201}, and is known as the gyrotropic magnetic effect. Note that these responses require inversion symmetry breaking.

Next, we discuss the transverse response to $\textbf{B}_{\omega}$, whose associated conductivity is given in Eq.~(\ref{eq18d}). Such contribution requires inversion-symmetry breaking and originates from the Fermi sea. In particular, it highly depends on the momentum space texture of magnetic moment of Bloch electrons, e.g., it vanishes in non-centrosymmetric materials displaying ``hedgehog"-like magnetic moment texture, favoring a Rashba-type configurations, such as the one in BiAg$_2$ monolayers~\cite{Go2017}.

In moving forward, we discuss the impact of $\textbf{E}_0$ on the AC transport coefficients, i.e., electro-optic effects. Here, $\textbf{E}_0$ is shown to impact the final AC current responses, giving rise to correlations between static and dynamical fields through the Berry curvature and the magnetic moment of Bloch electrons.

\begin{table}[]
    \caption{Allowed electro-Optic effects under time-reversal ($\mathcal{T}$) and inversion ($\mathcal{P}$) symmetries. The columns indicate electro-optic effects derived from the Berry curvature ( $\boldsymbol{\Omega}_{n\textbf{k}}$), magnetic moment ($\textbf{m}_{n\textbf{k}}$) or the simultaneous presence of both ($\boldsymbol{\Omega}_{n\textbf{k}} \& \textbf{m}_{n\textbf{k}}$). Rows indicate the type of electro-optic effect determined from its dependence on the AC fields; $\textbf{E}_{\omega}$ for Electric Drude-like response, $\times\textbf{E}_{\omega}$ for Electric Hall response, $\textbf{B}_{\omega}$ for magnetoelectric response, $\times\textbf{B}_{\omega}$ for transverse magnetoelectric response.}
    \centering
    \begin{tabular}{cccc}
        \hline\hline
         &  \ \ \ \ \ \  $\boldsymbol{\Omega}_{n\textbf{k}}$ \ \  \ \  \ \ \ \ &  \ \  \ \ \ \ $\textbf{m}_{n\textbf{k}}$ \ \ \ \ \ \ \ \ &  \ \ \ \  \ \ \ \ $\boldsymbol{\Omega}_{n\textbf{k}} \& \textbf{m}_{n\textbf{k}}$  \ \  \ \ \ \  \\
         \hline 
        $\ \ \textbf{E}_{\omega}$ & $\mathcal{T}$\cmark -- $\mathcal{P}$\xmark & --  & --   \\
       $\times \textbf{E}_{\omega}$ &  $\mathcal{T}$\cmark -- $\mathcal{P}$\xmark & --  & --   \\
        $\ \ \textbf{B}_{\omega}$ &  $\mathcal{T}$\xmark -- $\mathcal{P}$\cmark & $\mathcal{T}$\xmark -- $\mathcal{P}$\cmark &  $\mathcal{T}$\cmark -- $\mathcal{P}$\xmark \ \ \ding{120} $\mathcal{T}$\xmark -- $\mathcal{P}$\cmark \\
        $\times \textbf{B}_{\omega}$ & -- & $\mathcal{T}$\xmark -- $\mathcal{P}$\cmark &  -- \\
        \hline\hline
    \end{tabular}
    \label{tableI}
\end{table}

\section{V. Electro-Optic Effects}\label{sectionV}

We devote this section to study specific contributions to the AC response arising from the static $\textbf{E}_0$ field, which we refer to as $\textbf{J}^{\textbf{E}_0}_{\omega}$. For clarity, we separate the discussion of effects originating solely from Berry curvature, $\textbf{J}^{\textbf{E}_0\boldsymbol{\Omega}}_{\omega}$, and magnetic moment, $\textbf{J}^{\textbf{E}_0\textbf{m}}_{\omega}$, and those arising from the simultaneous presence of both quantities, $\textbf{J}^{\textbf{E}_0\boldsymbol{\Omega}\textbf{m}}_{\omega}$. As such, the final current response should be understood as the sum of the various contributions, i.e., $\textbf{J}^{\textbf{E}_0}_{\omega} = \textbf{J}^{\textbf{E}_0\boldsymbol{\Omega}}_{\omega} + \textbf{J}^{\textbf{E}_0\textbf{m}}_{\omega} + \textbf{J}^{\textbf{E}_0\boldsymbol{\Omega}\textbf{m}}_{\omega}$. We find that linear electro-optic effects comprise Drude-like responses ($\propto\textbf{E}_{\omega}$), electric Hall responses ($\propto\times\textbf{E}_{\omega}$), and their magnetoelectric counterparts, i.e., conventional magnetoelectric ($\propto\textbf{B}_{\omega}$) and transverse magnetoelectric ($\propto\times\textbf{B}_{\omega}$) responses. Table~\ref{tableI} summarizes all time-reversal ($\mathcal{T}$) and inversion ($\mathcal{P}$) symmetry allowed responses, which we describe in more detail in the following.

\subsection{A. Berry Curvature Contributions}

To start with, we describe the electro-optic effects derived from the Berry curvature of Bloch electrons. Our results indicate that the most general response can be summarized as
\begin{eqnarray}
J^{\textbf{E}_0\boldsymbol{\Omega}, \alpha}_{\omega} = \sigma^{\textbf{E}_0\boldsymbol{\Omega},\alpha\beta}_{\omega E}E^{\beta}_{\omega} + \epsilon_{\gamma\beta\alpha}\sigma^{\textbf{E}_0\boldsymbol{\Omega},\gamma}_{\omega HE}E^{\beta}_{\omega} +  \sigma^{\textbf{E}_0\boldsymbol{\Omega},\alpha\beta}_{\omega B} B^{\beta}_{\omega}, \nonumber \\
    \label{EO_J}
\end{eqnarray}
with transport coefficients explicitly given by
 \begin{subequations}
    \begin{eqnarray}
    \sigma^{\textbf{E}_0\boldsymbol{\Omega},\alpha\beta}_{\omega E} = \frac{e^3\tau/\hbar}{1-i\omega\tau}\sum_{n\textbf{k}} \left(-\frac{\partial f^0_{n\textbf{k}}}{\partial \epsilon_{n\textbf{k}}}\right)  (\textbf{E}_0 \times \boldsymbol{\Omega}_{n\textbf{k}})^{\alpha}v^{\beta}_{n\textbf{k}}, \nonumber \\
        \label{eq22a}
    \end{eqnarray}
        \begin{eqnarray}
    \boldsymbol{\sigma}^{\textbf{E}_0\boldsymbol{\Omega}}_{\omega HE} = \displaystyle  \frac{e^3\tau}{\hbar}\sum_{n\textbf{k}} \left(-\frac{\partial f^0_{n\textbf{k}}}{\partial \epsilon_{n\textbf{k}}}\right)  (\textbf{E}_0 \cdot \textbf{v}_{n\textbf{k}})\boldsymbol{\Omega}_{n\textbf{k}},
        \label{eq22b}
    \end{eqnarray}
        \begin{eqnarray}
    &\sigma_{\omega B}^{\textbf{E}_0\boldsymbol{\Omega},\alpha\beta} =  \displaystyle  \frac{e^3\tau}{\hbar}\sum_{n\textbf{k}} \left(-\frac{\partial f^0_{n\textbf{k}}}{\partial \epsilon_{n\textbf{k}}}\right)  (\textbf{E}_0 \cdot \textbf{v}_{n\textbf{k}}) (\textbf{v}_{n\textbf{k}} \cdot \boldsymbol{\Omega}_{n\textbf{k}}) \delta_{\alpha\beta},\nonumber \\ 
    & \displaystyle - \epsilon_{\lambda \gamma \beta} \frac{e^3\tau/\hbar}{1-i\omega\tau}\sum_{n\textbf{k}} \left(-\frac{\partial f^0_{n\textbf{k}}}{\partial \epsilon_{n\textbf{k}}}\right) v^{\alpha}_{n\textbf{k}}  (\textbf{E}_0 \times \boldsymbol{\Omega}_{n\textbf{k}})^{\lambda}v^{\gamma}_{n\textbf{k}}. \nonumber \\
        \label{eq22c}
    \end{eqnarray}
\end{subequations}   

Equations~(\ref{eq22a}) and (\ref{eq22b}) were recently discussed in Refs.~\cite{PhysRevLett.130.076901,PhysRevB.107.125151}, and describe non-conservative and conservative gyrotropic responses, respectively. While $\boldsymbol{\sigma}^{\textbf{E}_0\boldsymbol{\Omega}}_{\omega HE}$ is connected to an optical Hall effect, the frequency-dependent contribution $\sigma^{\textbf{E}_0\boldsymbol{\Omega},\alpha\beta}_{\omega E}$ has been shown to give rise to optical gain in non-centrosymmetric systems in recent works~\cite{PhysRevLett.130.076901, https://doi.org/10.48550/arxiv.2401.13764, https://doi.org/10.48550/arxiv.2312.15142}. These contributions are most commonly expressed in terms of the BCD tensor, with components $D^{\alpha\beta} = \sum_{n\textbf{k}} f_{n\textbf{k}}^0 C^{\alpha\beta}_{n\textbf{k}}$, where $C^{\alpha\beta}_{n\textbf{k}}= \nabla_{\textbf{k}}^{\beta}\Omega_{n\textbf{k}}^{\alpha}$ is the Jacobian of the Berry curvature vector field. Explicitly, $ \boldsymbol{\sigma}^{\textbf{E}_0\boldsymbol{\Omega}}_{\omega E} = -(e^3\tau/\hbar^2) \textbf{F}_0 \cdot \textbf{D}/(1 - i\omega\tau)$, where $F_0^{\alpha\lambda} = -\epsilon_{\alpha\lambda\gamma}E_0^{\gamma}$ is an antisymmetric tensor formed from the static electric field, and $\boldsymbol{\sigma}^{\textbf{E}_0\boldsymbol{\Omega}}_{\omega HE} = (e^3\tau/\hbar^2) \textbf{D}\cdot\textbf{E}_0$. Because $\boldsymbol{\sigma}^{\textbf{E}_0\boldsymbol{\Omega}}_{\omega E}$ is a tensor and typically $\boldsymbol{\sigma}^{\textbf{E}_0\boldsymbol{\Omega}}_{\omega E} \neq (\boldsymbol{\sigma}^{\textbf{E}_0\boldsymbol{\Omega}}_{\omega E})^T$ in 2D materials (where $\boldsymbol{\Omega}_{n\textbf{k}} = \Omega_{n\textbf{k}}^z\hat{\textbf{z}} \rightarrow \textbf{D}^T \neq -\textbf{D}$), such a non-conservative contribution is often referred to as a non-Hermitian electro-optic effect~\cite{PhysRevLett.130.076901, https://doi.org/10.48550/arxiv.2401.13764} ($\boldsymbol{\sigma}^{\textbf{E}_0\boldsymbol{\Omega}}_{\omega HE}$ is commonly referred to as a Hermitian contribution to the electro-optic effect). Finite linear-in-$\textbf{E}_{\omega}$ optical-Hall effects are constrained to vanish in inversion symmetric systems. Hence, these optical responses are expected to play an important role in non-centrosymmetric systems, such as chiral three-dimensional materials or twisted atomic bilayers~\cite{PhysRevLett.130.076901, https://doi.org/10.48550/arxiv.2401.13764}.

Besides the two electro-optic effect contributions mentioned above, our analysis reveals an extra contribution arising from the Berry curvature. The associated transport coefficient is given in Eq.~(\ref{eq22c}), which requires time-reversal symmetry breaking. The two contributions to $\sigma_{\omega B}^{\textbf{E}_0,\alpha\beta}$ describe an electro-optic effect that couples to the magnetic field sector of light and, thus, are associated with a magnetoelectric effect. The first contribution, longitudinal to $\textbf{B}_{\omega}$, originates from the $(\textbf{v}_{n\textbf{k}} \cdot \boldsymbol{\Omega}_{n\textbf{k}})$ term in the definition of the charge current, Eq.~(\ref{eq003}), and is the Fermi surface analogue of the Fermi sea bias-independent contribution given in Eq.~(\ref{eq18c}). In fact, the electro-optic effect in question can be obtained from the Fermi sea bias-independent response with the prescription $f^0_{n\textbf{k}} \rightarrow g^{E_0}_{n\textbf{k}}$. Therefore, we conclude that the requirements to observe such an electro-optic effect are analogous to the requirements needed to observe a chiral magnetic effect response, i.e., $\textbf{E}_0 \cdot \textbf{B} \neq 0$ in Weyl semimetals. For this reason, such contribution will be referred to as the \textit{Chiral-Magnetic Electro-Optic} effect from now on. The second contribution derives from the non-equilibrium  distribution function term given in Eq.~(\ref{eq88}), from where its frequency dependence originates. Recall that Eq.~(\ref{eq88}) carries the phase space volume correction due to the simultaneous presence of a Berry curvature and a magnetic field. In case such correction can safely be neglected, it follows that $\textbf{B}_{\omega}\times (\boldsymbol{\Omega}_{n\textbf{k}} \times \textbf{E}_0) \approx (\textbf{E}_0\cdot\textbf{B}_{\omega})\boldsymbol{\Omega}_{n\textbf{k}}$. Hence, the $\textbf{k}$-dependent contribution can be expressed as $(e^3\tau/\hbar) (\partial f^0_{n\textbf{k}}/\partial\epsilon_{n\textbf{k}}) \textbf{v}_{n\textbf{k}}(\textbf{v}_{n\textbf{k}} \cdot \boldsymbol{\Omega}_{n\textbf{k}})(\textbf{E}_0 \cdot \textbf{B}_{\omega})/(1 - i\omega\tau)$, which can be obtained from Eq.~(\ref{eq18c}) with the prescription, $f^0_{n\textbf{k}} \rightarrow (e^3\tau/\hbar) (\partial f^0_{n\textbf{k}}/\partial\epsilon_{n\textbf{k}}) \textbf{v}_{n\textbf{k}}(\textbf{E}_0 \cdot \textbf{B}_{\omega})/(1 - i\omega\tau)$. Therefore, we conclude that such contribution is also a type of Chiral-Magnetic Electro-Optic effect.

The above responses account for all possible linear electro-optic effects originating solely from the Berry curvature of Bloch electrons. The distinct contributions and the symmetry requirements for their existence are summarized in the first column of Table~\ref{tableI}. Next, we address the contribution originating solely from the magnetic moment of Bloch electrons.

\subsection{B. Magnetic Moment Contributions}

We find that electro-optic effects deriving solely from the magnetic moment of Bloch electrons can be summarized in the following constitutive relation
\begin{eqnarray}
J^{\textbf{E}_0\textbf{m}, \alpha}_{\omega} = \sigma^{\textbf{E}_0\textbf{m},\alpha\beta}_{\omega B} B^{\beta}_{\omega} + \epsilon_{\gamma\beta\alpha}\sigma^{\textbf{E}_0\textbf{m},\gamma}_{\omega HB}B^{\beta}_{\omega}, \nonumber \\
    \label{eq23}
\end{eqnarray}
with transport coefficients explicitly given by

\begin{subequations}    
\begin{eqnarray}
    &\sigma_{\omega B}^{\textbf{E}_0\textbf{m},\alpha\beta} =  \displaystyle \frac{e^2}{\hbar}\frac{1}{1 - i\omega\tau}\sum_{n\textbf{k}} \left(-\frac{\partial f^0_{n\textbf{k}}}{\partial \epsilon_{n\textbf{k}}}\right)  v^{\alpha}_{n\textbf{k}} K^{\beta\gamma}_{n\textbf{k}}E^{\gamma}_0, \nonumber \\
    &\displaystyle + \frac{e^2\tau}{\hbar} \sum_{n\textbf{k}} \left(\frac{\partial f^0_{n\textbf{k}}}{\partial \epsilon_{n\textbf{k}}}\right) (\textbf{E}_0 \cdot \textbf{v}_{n\textbf{k}}) K^{\alpha \beta}_{n\textbf{k}},
        \label{eq24a}
\end{eqnarray}
\begin{eqnarray}
    \boldsymbol{\sigma}^{\textbf{E}_0\textbf{m}}_{\omega HB} = \displaystyle  \frac{e^2\tau}{\hbar} \sum_{n\textbf{k}} \left(\frac{\partial f^0_{n\textbf{k}}}{\partial \epsilon_{n\textbf{k}}}\right) (\textbf{E}_0 \cdot \textbf{v}_{n\textbf{k}}) (\nabla_{\textbf{k}} \times \textbf{m}_{n\textbf{k}}), \nonumber \\
        \label{eq24b}
    \end{eqnarray}
\end{subequations}

Note that the above responses do not couple to the AC electric field and, thus, are generally magnetoelectric in nature. Both responses are only allowed in systems lacking time-reversal symmetry and are highly contingent on the magnetic moment texture of Bloch electrons in the $\textbf{k}$ space. 

The magnetoelectric response is divided into two main contributions, explicitly given in Eq.~(\ref{eq24a}). The first one arises from the correction to the non-equilibrium distribution function due to the magnetic moment of Bloch electrons given by Eq.~(\ref{eq8}); It exhibits a Drude-like frequency dependence and hinges on the magnetic moment texture of Bloch electrons through its Jacobian $\textbf{K}_{n\textbf{k}}$. The second contribution is frequency-independent and derives from the combination of the Zeeman-induced band velocity correction to the current integrand, Eq.~(\ref{eq3}), and the corrections to the non-equilibrium distribution function due to $\textbf{E}_0$. Its dependence on the components of $\textbf{K}_{n\textbf{k}}$ also imply in its magnetic moment texture-dependence.  

The transverse magnetoelectric response is frequency-independent and is the Fermi surface analogue of the static electric field-free case; It can be obtained from Eq.~(\ref{eq18d}) through the prescription $f^0_{n\textbf{k}} \rightarrow g^{E_0}_{n\textbf{k}}$. As such, it highly depends on the magnetic moment texture of Bloch electrons in momentum space and requires a Rashba-type texture to contribute to the final electro-optic effect. The two magnetoelectric responses discussed here are summarized in the second column of Table~\ref{tableI}, from where we conclude that it takes simultaneous time-reversal and inversion symmetry breaking and particular magnetic moment texture to generate an electro-optic effect from all possible AC field dependencies. 

So far we have focused on contributions to the electro-optic effect originating from either the Berry curvature or magnetic moment. We found that it is also possible to obtain responses that only exist if the Bloch states support both quantities simultaneously, which we discuss next. 

\subsection{C. Simultaneous Berry curvature and Magnetic Moment Contributions}

Finally, we address the contributions arising from the simultaneous presence of Berry curvature and magnetic moment of Bloch electrons. The associated current response is
\begin{eqnarray}
J^{\textbf{E}_0\boldsymbol{\Omega}\textbf{m}, \alpha}_{\omega} = \sigma^{\textbf{E}_0\boldsymbol{\Omega}\textbf{m},\alpha\beta}_{\omega B} B^{\beta}_{\omega}, \nonumber \\
    \label{eq23}
\end{eqnarray}
with a single transport coefficient given by
\begin{eqnarray}
    &\sigma_{\omega B}^{\textbf{E}_0\boldsymbol{\Omega}\textbf{m},\alpha\beta} =  \displaystyle \frac{e^2}{\hbar}\frac{i\omega\tau}{i\omega\tau - 1}\sum_{n\textbf{k}} \left(-\frac{\partial f^0_{n\textbf{k}}}{\partial \epsilon_{n\textbf{k}}}\right) (\textbf{E}_0 \times \boldsymbol{\Omega}_{n\textbf{k}})^{\alpha} m^{\beta}_{n\textbf{k}}. \nonumber \\
        \label{eq24}
\end{eqnarray}

\begin{figure*}[t]
\centerline{\includegraphics[width = 0.8\linewidth]{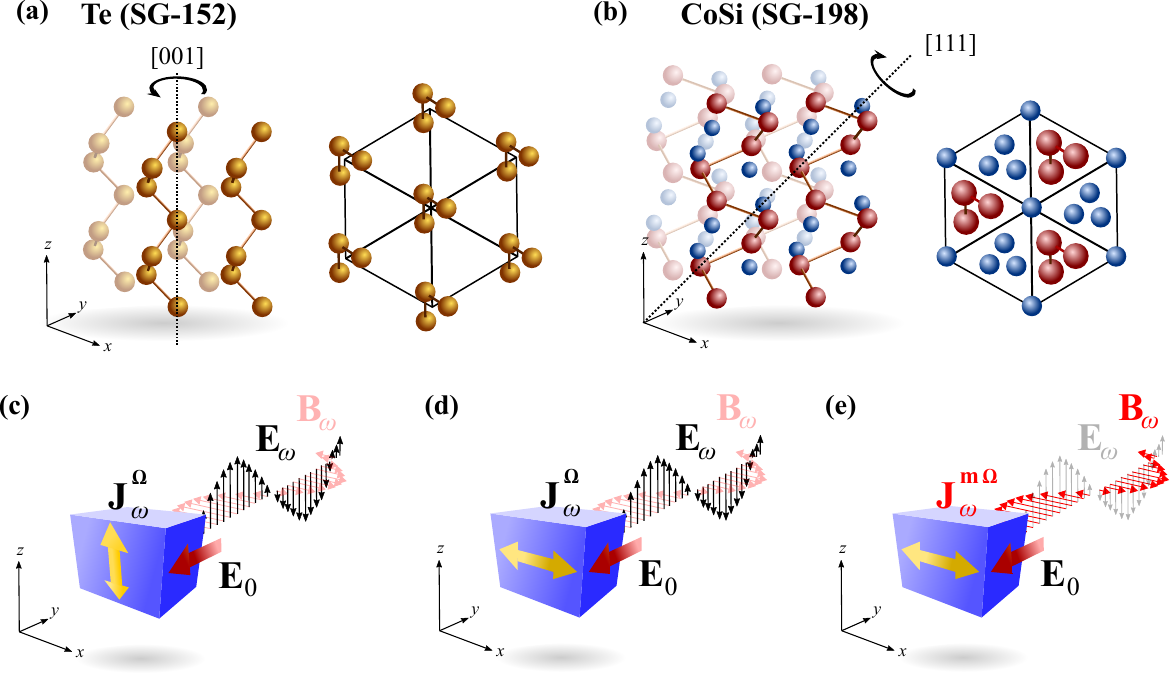}}
\caption{(a) Crystal structure of chiral Tellurium (SG-152). The chiral axis is represented by a dotted line, which coincides with the $z$ axis ($[001]$) in this case. The right panel highlights the crystal structure form the point of view of the chiral axis. (b) Crystal structure of chiral monosilicide crystals (SG-198). The chiral axis is now $[111]$, as represented by the dotted line. Crystal structure from the point of view of the chiral axis is also illustrated. (c)-(e) Linear current responses to $\textbf{E}_{\omega}$ and $\textbf{B}_{\omega}$ associated to electro-optic effects in non-centrosymmetric chiral quantum materials. Panels (c), (d) and (e) depict a gyrotropic electric response to $\textbf{E}_{\omega}$, a gyrotropic electric-Hall response to $\textbf{E}_{\omega}$ ($\propto \times \textbf{E}_{\omega}$) and a gyrotropic magnetoelectric response to $\textbf{B}_{\omega}$, respectively. The quantum mechanical properties of Bloch electrons responsible for these responses are highlighted in the superscript index in $\textbf{J}_{\omega}$, where $\boldsymbol{\Omega}$ and $\textbf{m}$ refer to the Berry curvature and magnetic moment of Bloch electrons, respectively.}
\label{Fig2}
\end{figure*}


The above contribution describes a magnetoelectric effect that derives from the interplay between the Zeeman-induced modification of the non-equilibrium distribution function of Bloch electrons, Eq.~(\ref{eq77}), and the anomalous Hall term appearing in the definition of the charge current, Eq.~(\ref{eq003}). This contribution describes a \textit{magnetoelectric electro-optic} effect and, thus, will be referred to by such terminology in this work. We note that such contribution is the magnetoelectric analogue of the non-Hermitian electro-optic effect given in Eq.~(\ref{eq22a}). However, Eq.~(\ref{eq24}) is not expressible as a BCD due to the absence of a band velocity $v^{\beta}_{n\textbf{k}}$ component in the integrand. In spite of that, it is possible to write this contribution in a compact form by introducing a new tensor $\textbf{G} = \sum_{n\textbf{k}}(-\partial f^0_{n\textbf{k}}/\partial \epsilon_{n\textbf{k}})\textbf{G}_{n\textbf{k}}$, dubbed \textit{magnetoelectric electro-optic tensor} in this work, whose components are related to $G^{\alpha\beta}_{n\textbf{k}} = \Omega^{\alpha}_{n\textbf{k}}m^{\beta}_{n\textbf{k}}$. The result is
\begin{eqnarray}
\sigma_{\omega B}^{\textbf{E}_0\boldsymbol{\Omega}\textbf{m},\alpha\beta}  = \displaystyle \frac{e^2}{\hbar}\frac{i\omega\tau}{i\omega\tau - 1} \textbf{F}_0 \cdot \textbf{G}.
    \label{eq25}
\end{eqnarray}

It is worth emphasizing that Eq.~(\ref{eq25}) is allowed in both non-centrosymmetric and broken time-reversal symmetric systems. This is due to the fact that Berry curvature and magnetic moment are constrained in the same manner by these symmetries, such that the components of $\textbf{G}_{n\textbf{k}}$ are even in $\textbf{k}$ in both cases. This is an unique feature not shared by the other electro-optic effects studied in this paper, that are allowed in \textit{either} non-centrosymmetric \textit{or} time-reversal broken systems, but not on both situations simultaneously as shown in Table~\ref{tableI}. In this sense, the magnetoelectric electro-optic effects are unavoidable when the system lacks either one of these symmetries. 

Following Ref.~\cite{PhysRevLett.116.077201}, we conclude that the magnetoelectric effect in biased time-reversal symmetric systems is described by the relations
\begin{subequations}
\begin{eqnarray}
  P_{\omega}^{\alpha} = \displaystyle \frac{i}{\omega} (\sigma_{\omega B}^{\alpha\beta} + \sigma_{\omega B}^{\textbf{E}_0\boldsymbol{\Omega}\textbf{m},\alpha\beta}) B_{\omega}^{\beta}  
\end{eqnarray}

\begin{eqnarray}
  M_{\omega}^{\beta} = \displaystyle -\frac{i}{\omega} (\sigma_{\omega B}^{\alpha\beta} + \sigma_{\omega B}^{\textbf{E}_0\boldsymbol{\Omega}\textbf{m},\alpha\beta}) E_{\omega}^{\alpha}  
\end{eqnarray}
\end{subequations}
for the polarization and magnetic moment induced by the oscillating electromagnetic field, with $J_{\omega}^{\alpha} = -i\omega P_{\omega}^{\alpha}$. That is, the magnetoelectric electro-optic contribution derived here describes the correction induced by $\textbf{E}_0$ to the magnetoelectric coupling coefficient $\sigma_{\omega B}^{\alpha\beta}$. 

In what follows, we apply the theory developed here to study electro-optic effects in non-centrosymmetric time-reversal symmetric systems. We choose bulk chiral materials from SG-152 and SG-198 as platforms to investigate the various electro-optical effects discussed in this work due to their well established texture of magnetic moment and Berry curvature~\cite{PhysRevB.97.035158, Yang2023}.  


\section{VI. Symmetry Imposed Shape of Electro-Optic Tensors of Chiral Materials}

We further extend the analysis of the previous sections to determine the transport coefficients associated with linear electro-optic effects in time-reversal symmetric chiral crystals. Here, we focus on materials pertaining to SG-152 and SG-198, such as chiral Te and CoSi, respectivelly, whose crystal structures are illustrated in Figs.~\ref{Fig2}(a) and (b). These materials have been shown to support pseudo-relativistic multifold fermions not accounted for by the standard model of particle physics~\cite{Chang2018, Bradlyn2016, PhysRevLett.130.066402, PhysRevLett.130.066402}. Such low energy quasiparticles are responsible for the various exotic charge and spin transport properties~\cite{Calavalle2022, PhysRevB.108.L201404}, spin textures~\cite{Chang2018, PhysRevLett.124.136404, PhysRevLett.125.216402}, as well as magnetoelectric and magneto-optical responses in these materials~\cite{Furukawa2017}. In this section, we aim to investigate how the SG of these particular material classes constrain the shape of the transport coefficients associated with electro-optic effects.

To start with, we summarize below the responses allowed by time-reversal symmetry describing linear electro-optic effects: 
\begin{eqnarray}
\textbf{J}^{\textrm{EO}}_{\omega} = \Bar{\boldsymbol{\sigma}}^{\textrm{EO}}_{\textrm{E}} \cdot \textbf{E}_{\omega} + \Bar{\boldsymbol{\sigma}}^{\textrm{EO}}_{\textrm{HE}} \times \textbf{E}_{\omega} + \Bar{\boldsymbol{\sigma}}^{\textrm{EO}}_{\textrm{B}} \cdot \textbf{B}_{\omega},
    \label{0eq26}
\end{eqnarray}
with the definitions
\begin{subequations}
\begin{eqnarray}
\Bar{\boldsymbol{\sigma}}^{\textrm{EO}}_{\textrm{E}}  = \displaystyle -\frac{e^3\tau}{\hbar}\frac{1}{1 - i\omega\tau} \textbf{F}_0 \cdot \textbf{D},
    \label{eq27a}
\end{eqnarray}
\begin{eqnarray}
\Bar{\boldsymbol{\sigma}}^{\textrm{EO}}_{\textrm{HE}} = \displaystyle \frac{e^3\tau}{\hbar} \textbf{D} \cdot \textbf{E}_0,
    \label{eq27b}
\end{eqnarray}
    \begin{eqnarray}
\Bar{\boldsymbol{\sigma}}^{\textrm{EO}}_{\textrm{B}}  = \displaystyle \frac{e^2}{\hbar}\frac{i\omega\tau}{i\omega\tau - 1} \textbf{F}_0 \cdot \textbf{G}.
    \label{eq27c}
\end{eqnarray}
\end{subequations}

The relevant tensors in the above equations have components $D^{\alpha\beta} = \sum_{n\textbf{k}}(-\partial f_{n\textbf{k}}^0/\partial\epsilon_{n\textbf{k}})\Omega_{n\textbf{k}}^{\alpha} v_{n\textbf{k}}^{\beta}$, corresponding to the BCD tensor, and  $G^{\alpha\beta} = \sum_{n\textbf{k}}(-\partial f_{n\textbf{k}}^0/\partial\epsilon_{n\textbf{k}})\Omega_{n\textbf{k}}^{\alpha} m_{n\textbf{k}}^{\beta}$, corresponding to the magnetoelectric electro-optic tensor. The DC electric field dependence is captured through either the antisymmetric $\textbf{F}_0$ tensor, with components $F_0^{\alpha\beta} = -\epsilon_{\alpha\beta\gamma}E_0^{\gamma}$, or the electric field vector $\textbf{E}_0 = [E_0^x \ \ E_0^y  \ \ E_0^z ]^T$, where the superscript stands for ``transpose". The responses described by Eq.~(\ref{0eq26}) are illustrated in Fig.~\ref{Fig2}(c)-(e); Equations~(\ref{eq27a}) and (\ref{eq27b}) describe responses to the oscillating electric field [panels (c) and (d)] originating from the BCD of Bloch electrons, whereas Eq.~(\ref{eq27c}) describes a response to the oscillating magnetic field, originating from the simultaneous presence of magnetic moment and Berry curvature of Bloch electrons. These responses are fully captured by the $\textbf{D}$ and $\textbf{G}$ tensors, which we describe next in more detail.

\subsection{A. Chiral Tellurium: SG-152}\label{sectionVIA}

In this section, we focus on trigonal Te~\cite{PhysRevB.97.205206,PhysRevB.97.035158, PhysRevLett.124.136404,PhysRevLett.125.216402, PhysRevB.99.245153,Furukawa2017}. Te belongs to a class of materials lacking mirror, inversion, and roto-inversion symmetries, namely chiral crystals~\cite{Chang2018}. Bulk Te is a non-magnetic, small-gap semiconductor that crystallizes in either of two enantiomorphic configurations~\cite{Qiu2022}. The atomic positions in both right-handed [SG-152] and left-handed [SG-154] enantiomorphs, are arranged in a helical manner along Te's three-fold trigonal axis, as shown in Fig.~\ref{Fig2}(a). 

We begin by studying the linear electro-optic effect tensor shapes imposed by the crystalline symmetries of Te. Because the electric and magnetic fields considered in this work are spatially uniform, only point group symmetries need to be considered when determining the transformation properties of the Berry curvature and total magnetic moment of a Bloch electrons. The group generators for Te consist of a three-fold rotation with respect to the trigonal axis and a two-fold rotation with respect to an axis in a perpendicular plane (see Appendix V). To simplify our symmetry analysis, we work in a cartesian coordinate system. These symmetries require that $(\Omega^{x}_{n(k_x,k_y,k_z)},\Omega^{y}_{n(k_x,k_y,k_z)}, \Omega^{z}_{n(k_x,k_y,k_z)}) = (\Omega^{x}_{n(k_x,-k_y,-k_z)},-\Omega^{y}_{n(k_x,-k_x,-k_z)}, -\Omega^{z}_{n(k_x,-k_y,-k_z)})$ and $(\Omega^{x}_{n(k_x,k_y,k_z)},\Omega^{y}_{n(k_x,k_y,k_z)}, \Omega^{z}_{n(k_x,k_y,k_z)}) = (-\frac{1}{2} \Omega^{x}_{n\textbf{k}^{'}} - \frac{\sqrt{3}}{2}  \Omega^{y}_{n\textbf{k}^{'}}, \frac{\sqrt{3}}{2}  \Omega^{x}_{n\textbf{k}^{'}} - \frac{1}{2}  \Omega^{y}_{n\textbf{k}^{'}}, \Omega^{z}_{n\textbf{k}^{'}})$, where $\textbf{k}^{'}=(-\frac{1}{2} k_x - \frac{\sqrt{3}}{2} k_y, \frac{\sqrt{3}}{2} k_x - \frac{1}{2} k_y, k_z)$. Identical constraints are imposed on the total magnetic moment.\\\\

First, we focus on the BCD tensor. Under the crystalline symmetries of Te, the BCD assumes the following form
\begin{equation}
\textbf{D} = 
\begin{bmatrix}
D & 0 & 0\\
0 & D & 0\\
0 & 0 & D^{zz}
\end{bmatrix},
\end{equation}
with $D^{xx} = D^{yy} = D$. Further, the locally divergence-free Berry curvature away from any Weyl points in Te requires the BCD tensor to be traceless, such that $D^{zz} = -2D$~\cite{PhysRevB.97.035158}. We note that, because right-handed and left-handed enantiomorphs of Te are related by an inversion operation, their BCDs are related by an overall minus sign.

The magnetoelectric electro-optic effect, which is described by Eq.~(\ref{eq27c}), is expressible in terms of the \textbf{G} tensor and originates from a coupling between Berry curvature and total magnetic moment of Bloch electrons. We again would like to emphasize that such a contribution can be present in systems with broken inversion symmetry or broken time-reversal symmetry, which is in contrast to all other electro-optic transport coefficients discussed in this work. The \textbf{G} tensor assumes the following shape
\begin{equation}
\textbf{G} = 
\begin{bmatrix}
G & 0 & 0\\
0 & G & 0\\
0 & 0 & G^{zz}
\end{bmatrix},
\end{equation}
with $G^{xx} = G^{yy} = G$, structurally similar to the BCD tensor of Te. However, the magnetoelectric electro-optic tensor need not be traceless and is identical in magnitude and sign for inequivalent enantiomorphs. This feature traces back to the properties of the Berry curvature and magnetic moment of Bloch electrons, which transform in the same way under all symmetry operations.  

Next, we address the symmetry imposed shape of electro-optic tensors of SG-198 materials. 

\subsection{B. Chiral Monosilicides: SG-198}\label{sectionVII}

In this section, we enforce the crystalline symmetries of SG-198 to determine the shape of the electro-optic tensors of topological chiral monosilicides, e.g., CoSi. These materials are structurally chiral, as illustrated in Fig.~\ref{Fig2}(b), with the helical arrangements of atomic positions along the [111] direction~\cite{Chang2018}. They belong to a specific non-symmorphic chiral SG that has been shown to support the largest possible Chern number that can be carried by a point-like band crossing~\cite{Chang2018, Schrter2019, PhysRevLett.130.066402}. Such large Chern numbers have been predicted to give rise to enhanced electron responses, such as spin Hall, and optical responses~\cite{PhysRevB.98.155145, PhysRevB.108.L201404}. 

Under the SG-198 symmetries, the BCD assumes the following shape
\begin{equation}
\textbf{D} = 
\begin{bmatrix}
D^{xx} & 0 & 0\\
0 & D^{yy} & 0\\
0 & 0 & D^{zz}
\end{bmatrix},
\end{equation}
Similar to the case of Te, the SG-198 BCD is diagonal. However, now $D^{xx} \neq D^{yy}$ in general due to the lack of additional symmetry constraints. The BCD of chiral monosilicides must also be traceless away from any point sources of Berry curvature, as required by a vanishing Berry curvature divergence, and must flip sign for the two enantiomorphs. 

The magnetoelectric electro-optic tensor for SG-198 is given by
\begin{equation}
\textbf{G} = 
\begin{bmatrix}
G^{xx} & 0 & 0\\
0 & G^{yy} & 0\\
0 & 0 & G^{zz}
\end{bmatrix}.
\end{equation}

This tensor is also diagonal, as a consequence of the symmetries of SG-198, but is not required to be traceless. Again, because the \textbf{G} tensor is invariant under an inversion operation, right-handed and left-handed CoSi have identical \textbf{G} tensors.

\section{VII. Conclusions}\label{sectionVIII}

In this work, we have investigated linear electro-optic effects due to Bloch electrons supporting concomitant Berry curvature and magnetic moment texture in momentum space. Our linear response theory is based on the semiclassical Boltzmann equation, and includes effects due to the presence of a static electric field $\textbf{E}_0$ and electromagnetic radiation, described through oscillating electric and magnetic fields $\textbf{E}_{\omega}$ and $\textbf{B}_{\omega}$. Our results indicated that the most general current response must consist of ``longitudinal" and transverse contributions to both $\textbf{E}_{\omega}$ and $\textbf{B}_{\omega}$ fields. The various contributions are shown to depend on time-reversal of inversion symmetry in quantum materials. In particular, electro-optic effects in time-reversal symmetric systems are shown to originate from the Berry curvature tensor, as previously discussed~\cite{PhysRevLett.130.076901, https://doi.org/10.48550/arxiv.2401.13764}, and from a new tensor, dubbed $\textbf{G}$, derived from the simultaneous presence of Berry curvature and magnetic moment of Bloch electrons. The latter contribution, which we refer to as magnetoelectric electro-optic effect, is shown to exist in either time-reversal symmetric or inversion symmetric systems and describes a response to the oscillating magnetic field of light, $\textbf{B}_{\omega}$, and static bias $\textbf{E}_0$. Finally, we have shown how the electro-optic tensor shapes are constrained by the symmetries of certain non-centrosymmetric chiral materials, pertaining to space groups 152 and 198. This work unveil unexplored pathways for tailoring electromagnetic responses of quantum materials, offering a rich playground for theoretical and experimental exploration.


\section{Acknowledgments} 
The authors acknowledge partial support  from Office of Naval Research MURI grant N00014-23-1-2567.

\begin{widetext}
    \section{Appendix I}\label{AppI}
Here, we provide further details regarding the semiclassical equations of motion. The equations are
\begin{eqnarray}
    \textbf{\.x}_{n\textbf{k}} = \textbf{v}_{n\textbf{k}} - \textbf{\.k}\times\boldsymbol{\Omega}_{n\textbf{k}},
    \label{eq01}
\end{eqnarray}
\begin{eqnarray}
    \hbar\textbf{\.k} = -e\textbf{E} - e \textbf{\.x}_{n\textbf{k}} \times \textbf{B},
    \label{eq02}
\end{eqnarray}
where $\boldsymbol{\Omega}_{n\textbf{k}}$ is the Berry curvature associated with the Bloch state, $|u_{n\textbf{k}}\rangle$ with energy $\epsilon_{n\textbf{k}}$, given by
\begin{eqnarray}
\boldsymbol{\Omega}_{n\textbf{k}} = -\operatorname{Im}\langle \nabla_{\textbf{k}} u_{n\textbf{k}}|\times |\nabla_{\textbf{k}}u_{n\textbf{k}}\rangle,
    \label{eq03}
\end{eqnarray}
and $\textbf{v}_{n\textbf{k}} = (1/\hbar)\partial \epsilon_{n\textbf{k}}/\partial \textbf{k}$ is the band velocity. The band velocity is related to the magnetic moment of Bloch electrons, $\textbf{m}_{n\textbf{k}}$, through the prescription $\hbar\textbf{v}_{n\textbf{k}} \rightarrow \hbar\textbf{v}_{n\textbf{k}} - \nabla_{\textbf{k}}(\textbf{m}_{n\textbf{k}}\cdot \textbf{B})$. Equations~(\ref{eq01}) and (\ref{eq02}) can be easily decoupled~\cite{PhysRevB.88.104412}. The procedure renders
\begin{subequations}
    \begin{eqnarray}
    \textbf{\.x}_{n\textbf{k}} = \left[1 + \frac{e}{\hbar}(\boldsymbol{\Omega}_{n\textbf{k}}\cdot \textbf{B})\right]^{-1} \left[\textbf{v}_{n\textbf{k}} + \left(\frac{e}{\hbar}\right)\textbf{E}\times\boldsymbol{\Omega}_{n\textbf{k}} + \left(\frac{e}{\hbar }\right)(\boldsymbol{\Omega}_{n\textbf{k}} \cdot \textbf{v}_{n\textbf{k}})\textbf{B}\right],
        \label{eqxx}
    \end{eqnarray}
    \begin{eqnarray}
    \textbf{\.k} = \left[1 + \frac{e}{\hbar}(\boldsymbol{\Omega}_{n\textbf{k}}\cdot \textbf{B})\right]^{-1}\left[ -\frac{e}{\hbar}\textbf{E} - \frac{e}{\hbar}\textbf{v}_{n\textbf{k}}\times\textbf{B} - \left(\frac{e}{\hbar}\right)^2 (\textbf{E}\cdot\textbf{B})\boldsymbol{\Omega}_{n\textbf{k}} \right],
        \label{kk}
    \end{eqnarray}
\end{subequations}
which are the main equations to be utilized in the Boltzmann formalism.

\section{Appendix II}\label{AppII}

Here, we address the derivation of the main Boltzmann equation utilized in the main text. The Boltzmann equation in its fullness is
\begin{eqnarray}
\frac{\partial }{\partial t}g_{n\textbf{k}} + \textbf{\.x}_{n\textbf{k}}\cdot \frac{\partial }{\partial \textbf{x}}g_{n\textbf{k}} + \textbf{\.k}\cdot \frac{\partial }{\partial \textbf{k}}g_{n\textbf{k}} = -\frac{\delta g_{n\textbf{k}}}{\tau}
    \label{boltzmanneq}
\end{eqnarray}
within the relaxation time approximation, characterized by $\tau$. The spatially homogeneous limit, i.e., $\partial g_{n\textbf{k}}/\partial \textbf{x} = 0$, is assumed throughout. Hence, the Boltzmann equation becomes
\begin{eqnarray}
    \frac{\partial }{\partial t}g_{n\textbf{k}} + \left[1 + \frac{e}{\hbar}(\boldsymbol{\Omega}_{n\textbf{k}}\cdot \textbf{B})\right]^{-1}\left[ -\frac{e}{\hbar}\textbf{E} - \frac{e}{\hbar}\textbf{v}_{n\textbf{k}}\times\textbf{B} - \left(\frac{e}{\hbar}\right)^2 (\textbf{E}\cdot\textbf{B})\boldsymbol{\Omega}_{n\textbf{k}} \right]\cdot \frac{\partial }{\partial \textbf{k}}g_{n\textbf{k}} = -\frac{\delta g_{n\textbf{k}}}{\tau},
    \label{eq04}
\end{eqnarray}    

One assumption in our formalism is that the simultaneous presence of $\boldsymbol{\Omega}_{n\textbf{k}}$ and $\textbf{B}$ are not enough to distort the phase-space volume considerably. This means that the Boltzmann equation is equivalent to

\begin{eqnarray}
    \frac{\partial }{\partial t}g_{n\textbf{k}} + \left[1 - \frac{e}{\hbar}(\boldsymbol{\Omega}_{n\textbf{k}}\cdot \textbf{B}) - \left(\frac{e}{\hbar}\right)^2(\boldsymbol{\Omega}_{n\textbf{k}}\cdot \textbf{B})^2 + \cdots\right]\left[ -\frac{e}{\hbar}\textbf{E} - \left(\frac{e}{\hbar}\right)^2 (\textbf{E}\cdot\textbf{B})\boldsymbol{\Omega}_{n\textbf{k}} \right]\cdot \hbar\textbf{v}_{n\textbf{k}}\frac{\partial f^0_{n\textbf{k}}}{\partial \epsilon_{n\textbf{k}}} = -\frac{\delta g_{n\textbf{k}}}{\tau},
    \label{eq04}
\end{eqnarray}  
after we expand the phase-space volume correction in powers of $(\boldsymbol{\Omega}_{n\textbf{k}}\cdot \textbf{B})$. Thus, in the small field limit, where third order terms in the fields are neglected, we get

\begin{eqnarray}
    \frac{\partial }{\partial t}g_{n\textbf{k}} + \left[1 - \frac{e}{\hbar}(\boldsymbol{\Omega}_{n\textbf{k}}\cdot \textbf{B}) \right]\left[ -\frac{e}{\hbar}\textbf{E} - \left(\frac{e}{\hbar}\right)^2 (\textbf{E}\cdot\textbf{B})\boldsymbol{\Omega}_{n\textbf{k}} \right]\cdot \hbar\textbf{v}_{n\textbf{k}}\frac{\partial f^0_{n\textbf{k}}}{\partial \epsilon_{n\textbf{k}}} = -\frac{\delta g_{n\textbf{k}}}{\tau},
    \label{eq04}
\end{eqnarray}    

or, more compactly
\begin{eqnarray}
    \frac{\partial }{\partial t}g_{n\textbf{k}} + \left[ -\frac{e}{\hbar}\textbf{E} - \left(\frac{e}{\hbar}\right)^2 (\textbf{E}\cdot\textbf{B})\boldsymbol{\Omega}_{n\textbf{k}} + \left(\frac{e}{\hbar}\right)^2  \textbf{E}(\boldsymbol{\Omega}_{n\textbf{k}}\cdot \textbf{B})\right]\cdot \hbar\textbf{v}_{n\textbf{k}}\frac{\partial f^0_{n\textbf{k}}}{\partial \epsilon_{n\textbf{k}}} = -\frac{\delta g_{n\textbf{k}}}{\tau}.
    \label{eq04}
\end{eqnarray}    
But, $\textbf{B}\times (\boldsymbol{\Omega}_{n\textbf{k}} \times \textbf{E}) = (\textbf{E}\cdot\textbf{B})\boldsymbol{\Omega}_{n\textbf{k}}  - \textbf{E}(\boldsymbol{\Omega}_{n\textbf{k}}\cdot \textbf{B})$. Therefore, the Boltzmann equation resumes to
\begin{eqnarray}
    \frac{\partial }{\partial t}g_{n\textbf{k}} + \left[ -\frac{e}{\hbar}\textbf{E} - \left(\frac{e}{\hbar}\right)^2 \textbf{B}\times (\boldsymbol{\Omega}_{n\textbf{k}} \times \textbf{E})\right]\cdot \hbar\textbf{v}_{n\textbf{k}}\frac{\partial f^0_{n\textbf{k}}}{\partial \epsilon_{n\textbf{k}}} = -\frac{\delta g_{n\textbf{k}}}{\tau}.
    \label{eq04}
\end{eqnarray}    

Next, we add the contribution arising from the magnetic moment of Bloch electrons through $\hbar\textbf{v}_{n\textbf{k}} \rightarrow \hbar\textbf{v}_{n\textbf{k}} - \nabla_{\textbf{k}}(\textbf{m}_{n\textbf{k}}\cdot \textbf{B})$. This procedure renders 
\begin{eqnarray}
    \frac{\partial }{\partial t}g_{n\textbf{k}} + \left[ -e\textbf{E}\cdot \textbf{v}_{n\textbf{k}} - \frac{e^2}{\hbar} [\textbf{B}\times (\boldsymbol{\Omega}_{n\textbf{k}} \times \textbf{E})]\cdot \textbf{v}_{n\textbf{k}} + \frac{e}{\hbar}\textbf{E}\cdot[\nabla_{\textbf{k}}(\textbf{m}_{n\textbf{k}}\cdot\textbf{B})]\right]\frac{\partial f^0_{n\textbf{k}}}{\partial \epsilon_{n\textbf{k}}} = -\frac{\delta g_{n\textbf{k}}}{\tau}.
    \label{eq04}
\end{eqnarray}   
In particular, the $\textbf{E}\cdot[\nabla_{\textbf{k}}(\textbf{m}_{n\textbf{k}}\cdot\textbf{B})]$ contribution can be rewritten in terms of the Jacobian of the vector field $\textbf{m}_{n\textbf{k}}$, as $\textbf{B}\cdot \textbf{K}_{n\textbf{k}}\cdot \textbf{E}$, where the Jacobian tensor, $ \textbf{K}_{n\textbf{k}}$, has components $ K^{\alpha\beta}_{n\textbf{k}} = \nabla^{\beta}_{\textbf{k}}m^{\alpha}_{n\textbf{k}}$. Here, $\textbf{K}_{n\textbf{k}}\cdot \textbf{E}$ is the 3D vector $\sum_{\beta}E^{\beta} \nabla^{\beta} \textbf{m}_{n\textbf{k}}$. Finally, we arrive at 
\begin{eqnarray}
    \frac{\partial }{\partial t}g_{n\textbf{k}} + \left[ -e\textbf{E}\cdot \textbf{v}_{n\textbf{k}} - \frac{e^2}{\hbar} [\textbf{B}\times (\boldsymbol{\Omega}_{n\textbf{k}} \times \textbf{E})]\cdot \textbf{v}_{n\textbf{k}} + \frac{e}{\hbar}\textbf{B}\cdot \textbf{K}_{n\textbf{k}}\cdot \textbf{E}\right]\frac{\partial f^0_{n\textbf{k}}}{\partial \epsilon_{n\textbf{k}}} = -\frac{\delta g_{n\textbf{k}}}{\tau}.
    \label{eq04}
\end{eqnarray} 
This is the main homogeneous Boltzmann equation utilized in this work.

\section{Appendix III}\label{AppIII}

In this appendix, we briefly indicate the derivation of the non-equilibrium distribution function. Solutions to the Boltzmann equation are assumed to be $g_{n\textbf{k}} = f^0_{n\textbf{k}} + \delta g_{n\textbf{k}} $, where $f^0_{n\textbf{k}}$ is the equilibrium Fermi-Dirac distribution of Bloch electrons and the non-equilibrium contribution, $\delta g_{n\textbf{k}} $, is
\begin{eqnarray}
    \delta g_{n\textbf{k}} = \delta g^{E_0}_{n\textbf{k}} + \delta g^{\omega}_{n\textbf{k}}e^{-i\omega t} + \delta g^{\omega *}_{n\textbf{k}}e^{i\omega t},
    \label{eq4}
\end{eqnarray}
 with static and dynamical contributions $g^{E_0}_{n\textbf{k}}$ and $\delta g^{\omega}_{n\textbf{k}}$, respectively. Plugging in this \textit{ansatz} into the Boltzmann equation renders the following relation

\begin{eqnarray}
    &\displaystyle \frac{\partial f^0_{n\textbf{k}}}{\partial t} - i\omega \delta g^{\omega}_{n\textbf{k}}e^{-i\omega t} + i\omega \delta g^{\omega*}_{n\textbf{k}}e^{i\omega t} + \nonumber \\ 
    &\displaystyle \left[ -e\textbf{E}\cdot \textbf{v}_{n\textbf{k}} - \frac{e^2}{\hbar} [\textbf{B}\times (\boldsymbol{\Omega}_{n\textbf{k}} \times \textbf{E})]\cdot \textbf{v}_{n\textbf{k}} + \frac{e}{\hbar}\textbf{B}\cdot \textbf{K}_{n\textbf{k}}\cdot \textbf{E}\right]\frac{\partial f^0_{n\textbf{k}}}{\partial \epsilon_{n\textbf{k}}} = -\frac{\delta g^{E_0}_{n\textbf{k}}}{\tau} -\frac{\delta g^{\omega}_{n\textbf{k}}}{\tau}e^{-i\omega t} -\frac{\delta g^{\omega *}_{n\textbf{k}}}{\tau}e^{i\omega t},
    \nonumber \\
    \label{eq04}
\end{eqnarray}
where $\textbf{E} = \textbf{E}_0 + \textbf{E}_{\omega}e^{-i\omega t} + \textbf{E}^*_{\omega}e^{i\omega t}$ and $\textbf{B} = \textbf{B}_{\omega}e^{-i\omega t} + \textbf{B}^*_{\omega}e^{i\omega t}$. Here, the time derivative of the fermi-dirac distribution is to be understood as
\begin{eqnarray}
    \frac{\partial f^0_{n\textbf{k}}}{\partial t} = \frac{\partial f^0_{n\textbf{k}}}{\partial \epsilon_{n\textbf{k}}}\frac{\partial \epsilon_{n\textbf{k}}}{\partial t} = - \frac{\partial f^0_{n\textbf{k}}}{\partial \epsilon_{n\textbf{k}}} \textbf{m}_{n\textbf{k}} \cdot \frac{\partial \textbf{B}}{\partial t} = i\omega  \frac{\partial f^0_{n\textbf{k}}}{\partial \epsilon_{n\textbf{k}}} \textbf{m}_{n\textbf{k}}\cdot \textbf{B}_{\omega}e^{-i\omega t} - i\omega  \frac{\partial f^0_{n\textbf{k}}}{\partial \epsilon_{n\textbf{k}}} \textbf{m}_{n\textbf{k}}\cdot \textbf{B}^{*}_{\omega}e^{i\omega t},
    \label{eq04}
\end{eqnarray}
as discussed in Ref.~\cite{PhysRevLett.116.077201}. The Boltzmann equation assumes the general form
\begin{eqnarray}
D^0_{n\textbf{k}} + D^{\omega}_{n\textbf{k}}e^{-i\omega t} + D^{\omega*}_{n\textbf{k}}e^{i\omega t} = 0,
    \label{eq15}
\end{eqnarray}
where we have defined
\begin{eqnarray}
D^0_{n\textbf{k}} = \displaystyle\frac{1}{\tau}\delta g^{E_0}_{n\textbf{k}} - e\textbf{E}_0\cdot \textbf{v}_{n\textbf{k}}\frac{\partial f^0_{n\textbf{k}}}{\partial \epsilon_{n\textbf{k}}} ,
    \label{eq16}
\end{eqnarray}
\begin{eqnarray}
& D^{\omega}_{n\textbf{k}} = \displaystyle\left(\frac{1}{\tau} - i\omega\right)\delta g^{\omega}_{n\textbf{k}} + i\omega \frac{\partial f^0_{n\textbf{k}}}{\partial \epsilon_{n\textbf{k}}} \textbf{m}_{n\textbf{k}}\cdot \textbf{B}_{\omega} + \left[-e\textbf{E}_{\omega} - \frac{e^2}{\hbar} \textbf{B}_{\omega}\times (\boldsymbol{\Omega}_{n\textbf{k}} \times \textbf{E}_0)\right]\cdot \textbf{v}_{n\textbf{k}}\frac{\partial f^0_{n\textbf{k}}}{\partial \epsilon_{n\textbf{k}}} + \nonumber \\
& \displaystyle \left(\frac{e}{\hbar}\right)\left[ \textbf{B}_{\omega}\cdot \textbf{K}_{n\textbf{k}}\cdot \textbf{E}_0\right]\frac{\partial f^0_{n\textbf{k}}}{\partial \epsilon_{n\textbf{k}}}
    \label{eq16}
\end{eqnarray}

Because the linear independence of the $e^{i\omega t}$s in Eq.~(\ref{eq15}) implies in $D^0_{n\textbf{k}} = 0$ and $D^{\omega}_{n\textbf{k}} = 0$, we arrive at  
\begin{eqnarray}
    \delta g^{E_0}_{n\textbf{k}} = \frac{\partial f^0_{n\textbf{k}}}{\partial \epsilon_{n\textbf{k}}} e \tau \textbf{v}_{n\textbf{k}} \cdot \textbf{E}_0,
    \label{eq5}
\end{eqnarray}
corresponding to the usual DC Drude response to $\textbf{E}_0$ and that the impact of the dynamical fields can most generally be separated into four distinct contributions, $\delta g^{\omega}_{n\textbf{k}} = \delta g^{E_{\omega}}_{n\textbf{k}} + \delta g^{B_{\omega}}_{n\textbf{k}} + \delta g^{\boldsymbol{\Omega}-E_0B_{\omega}}_{n\textbf{k}}  + \delta g^{\textbf{m}-E_0B_{\omega}}_{n\textbf{k}}$, which we list below:  

\begin{eqnarray}
    \delta g^{E_{\omega}}_{n\textbf{k}} = \frac{\partial f^0_{n\textbf{k}}}{\partial \epsilon_{n\textbf{k}}} \frac{e \tau}{1 - i\omega\tau} \textbf{v}_{n\textbf{k}} \cdot \textbf{E}_{\omega},
    \label{eq6}
\end{eqnarray}
corresponds to the typical AC Drude response to the oscillating electric field, $\textbf{E}_{\omega}$. The second AC contribution, 
\begin{eqnarray}
    \delta g^{B_{\omega}}_{n\textbf{k}} = \frac{\partial f^0_{n\textbf{k}}}{\partial \epsilon_{n\textbf{k}}} \frac{i\omega\tau}{i\omega\tau - 1} \textbf{m}_{n\textbf{k}} \cdot \textbf{B}_{\omega},
    \label{eq7}
\end{eqnarray}
describes how the AC magnetic field perturbs the distribution of Bloch electrons on the Fermi surface, by coupling to their magnetic moment. 

The third AC contribution to the non-equilibrium distribution function is
\begin{eqnarray}
    \delta g^{\boldsymbol{\Omega}-E_0B_{\omega}}_{n\textbf{k}} = \frac{\partial f^0_{n\textbf{k}}}{\partial \epsilon_{n\textbf{k}}} \frac{e^2 \tau/\hbar}{1 - i\omega\tau} [\textbf{B}_{\omega}\times (\boldsymbol{\Omega}_{n\textbf{k}} \times \textbf{E}_{0})]\cdot \textbf{v}_{n\textbf{k}},
    \label{eq8}
\end{eqnarray}
The fourth AC contribution to the non-equilibrium distribution function is
\begin{eqnarray}
    \delta g^{\textbf{m}-E_0B_{\omega}}_{n\textbf{k}} = -\frac{\partial f^0_{n\textbf{k}}}{\partial \epsilon_{n\textbf{k}}} \frac{e \tau/\hbar}{1 - i\omega\tau} [\textbf{B}_{\omega}\cdot \textbf{K}_{n\textbf{k}}\cdot \textbf{E}_0],
    \label{eq8}
\end{eqnarray}

Once established the solution to the Boltzmann equation, we proceed by deriving the constitutive relation.

\section{Appendix IV} \label{AppIV}

In this appendix, we write down the charge current expression consistent with our formalism. The charge current is
\begin{eqnarray}
    &\textbf{J} = -e\displaystyle\sum_{n\textbf{k}}  g_{n\textbf{k}} \left[\textbf{v}_{n\textbf{k}} + \left(\frac{e}{\hbar}\right)\textbf{E}\times\boldsymbol{\Omega}_{n\textbf{k}} + \left(\frac{e}{\hbar }\right)(\boldsymbol{\Omega}_{n\textbf{k}} \cdot \textbf{v}_{n\textbf{k}})\textbf{B}\right] + \nonumber \\
    &\displaystyle\frac{e}{\hbar}\sum_{n\textbf{k}}  g_{n\textbf{k}} \left[(\textbf{B}\cdot\nabla_{\textbf{k}})\textbf{m}_{n\textbf{k}} + \textbf{B}\times(\nabla_{\textbf{k}}\times\textbf{m}_{n\textbf{k}}) + \frac{e}{\hbar} (\textbf{B}\cdot \textbf{K}_{n\textbf{k}}\cdot \boldsymbol{\Omega}_{n\textbf{k}})\textbf{B} \right]. \label{eq3}
\end{eqnarray}

By replacing the full non-equilibrium distribution function derived previous, one obtains the main current response presented in the main text including all electro-optic effects discussed.

\section{Appendix V} 

Here, we briefly recap the constraints imposed over the Berry curvature and magnetic moment of Bloch electrons by time-reversal and inversion symmetries. To account for the symmetry transformation properties of both the Berry curvature and the total magnetic moment of a Bloch electron, we make the following substitution into equation (28), $\epsilon_{n\textbf{k}} \rightarrow \epsilon_{n\textbf{k}} - \textbf{m}_{n\textbf{k}} \cdot \textbf{B}$. Combining with (29), the semiclassical equation of motion (SEOM) becomes

\begin{eqnarray}
    \textbf{\.x}_{n\textbf{k}} = \frac{1}{\hbar} \grad_{\textbf{k}} \epsilon_{n\textbf{k}}  - \frac{1}{\hbar} \grad_{\textbf{k}}(\textbf{m}_{n\textbf{k}} \cdot \textbf{B}) +\frac{e}{\hbar}\textbf{E}\times\boldsymbol{\Omega}_{n\textbf{k}} +\frac{e}{\hbar}\textbf{\.x}_{n\textbf{k}} \times \textbf{B}\times\boldsymbol{\Omega}_{n\textbf{k}},
    \label{}
\end{eqnarray}

Under time-reversal the fields and dynamical variables transform as $\textbf{E} \rightarrow \textbf{E}$, $\textbf{B} \rightarrow -\textbf{B}$, $\dot{\textbf{x}}_{n\textbf{k}} \rightarrow -\dot{\textbf{x}}_{n(-\textbf{k})}$, and $\grad_{\textbf{k}} \epsilon_{n\textbf{k}} \rightarrow -\grad_{\textbf{k}} \epsilon_{n(-\textbf{k})}$. Under spatial inversion: $\textbf{E} \rightarrow -\textbf{E}$, $\textbf{B} \rightarrow \textbf{B}$, $\dot{\textbf{x}}_{n\textbf{k}} \rightarrow -\dot{\textbf{x}}_{n(-\textbf{k})}$ , and $\grad_{\textbf{k}} \epsilon_{n\textbf{k}} \rightarrow -\grad_{\textbf{k}} \epsilon_{n(-\textbf{k})}$.\\

Under a time-reversal transformation the SEOM becomes
\begin{eqnarray}
    -\textbf{\.x}_{n(-\textbf{k})} = -\frac{1}{\hbar}\grad_{\textbf{k}} \epsilon_{n(-\textbf{k})} + \frac{1}{\hbar} \grad_{\textbf{k}}(\textbf{m}_{n(-\textbf{k})} \cdot (-\textbf{B}))  + \frac{e}{\hbar}\textbf{E}\times\boldsymbol{\Omega}_{n(-\textbf{k})} +\frac{e}{\hbar}(-\textbf{\.x}_{n(-\textbf{k})}) \times (-\textbf{B})\times\boldsymbol{\Omega}_{n(-\textbf{k})},
    \label{}
\end{eqnarray}

However, with time-reversal symmetry (TRS) present $\epsilon_{n(-\textbf{k})} = \epsilon_{n\textbf{k}}$. Therefore, the covariance of the SEOM under time-reversal requires that $\boldsymbol{\Omega}_{n(-\textbf{k})} = -\boldsymbol{\Omega}_{n\textbf{k}}$ and  $\boldsymbol{m}_{n(-\textbf{k})} = -\boldsymbol{m}_{n\textbf{k}}$ , i.e. the Berry curvature and total magnetic moment must be odd functions of the crystal momentum for systems with TRS. Thus, properties of systems with TRS involving the sum of the Berry curvature or total moment over the full Brillouin zone vanish, e.g. $\sum_{\textbf{k} \in BZ}\boldsymbol{\Omega}_{n\textbf{k}} = \frac{1}{2}\sum_{\textbf{k} \in BZ}[\boldsymbol{\Omega}_{n\textbf{k}} + \boldsymbol{\Omega}_{n(-\textbf{k})}] = \frac{1}{2}\sum_{\textbf{k} \in BZ}[\boldsymbol{\Omega}_{n\textbf{k}} - \boldsymbol{\Omega}_{n\textbf{k}}] = 0$. \\

Under spatial inversion

\begin{eqnarray}
    -\textbf{\.x}_{n(-\textbf{k})} = -\frac{1}{\hbar}\grad_{\textbf{k}} \epsilon_{n(-\textbf{k})} + \frac{1}{\hbar} \grad_{\textbf{k}}(\textbf{m}_{n(-\textbf{k})} \cdot \textbf{B}) + \frac{e}{\hbar}(-\textbf{E})\times\boldsymbol{\Omega}_{n(-\textbf{k})} +\frac{e}{\hbar}(-\textbf{\.x}_{n(-\textbf{k})}) \times \textbf{B}\times\boldsymbol{\Omega}_{n(-\textbf{k})},
    \label{}
\end{eqnarray}

Spatial inversion symmetry implies $\epsilon_{n(-\textbf{k})} = \epsilon_{n\textbf{k}}$. The covariance of the SEOM under inversion requires that $\boldsymbol{\Omega}_{n(-\textbf{k})} = \boldsymbol{\Omega}_{n\textbf{k}}$ and $\boldsymbol{m}_{n(-\textbf{k})} = \boldsymbol{m}_{n\textbf{k}}$. If both time-reversal and inversion symmetries are present, then $\boldsymbol{\Omega}_{n\textbf{k}} = -\boldsymbol{\Omega}_{n(-\textbf{k})} = -\boldsymbol{\Omega}_{n\textbf{k}} = 0$ and $\boldsymbol{m}_{n\textbf{k}} = -\boldsymbol{m}_{n(-\textbf{k})} = -\boldsymbol{m}_{n\textbf{k}} = 0$. Thus, space-time inversion symmetry enforces the vanishing of the Berry curvature and total moment throughout the Brillouin zone. \\

To determine how the Berry curvature and total moment transform under the crystalline symmetries of right-handed and left-handed Tellurium [space group 152 (SG-152) and space group 154 (SG-154), respectively] we adopt the following set of group generators. SG-152: $\{C_{3_{001}}|0 0 \frac{1}{3}\}$ and $\{C_{2_{110}} | 0 0 0\}$, SG-154: $\{C_{3_{001}}|0 0 \frac{2}{3}\}$ and $\{C_{2_{110}} | 0 0 0\}.$ We note that real-space translations have no effect in reciprocal space. Additionally, because the dynamical variables do not depend on the real-space position and the fields considered are spatially uniform, the transformation properties of the Berry curvature and total moment depend only on the point group symmetries of SG-152 and SG-154, i.e. point group 32. In terms of their Cartesian components, the unit vectors along a,b, and c(trigonal) axes are $\hat{a} = \frac{1}{2} \hat{x} - \frac{\sqrt{3}}{2} \hat{y}$, $\hat{b} = \frac{1}{2} \hat{x} + \frac{\sqrt{3}}{2} \hat{y}$, and $\hat{c} = \hat{z}$, respectively. To simplify our symmetry analysis, we work in the Cartesian coordinate system.\\

We note that both the Berry curvature and the total moment are pseudovectors, thus must transform as such under the point group symmetry operators. In the Cartesian coordinate system the two-fold and three-fold rotations are represented by $R_{2x} = \begin{bmatrix}
1 & 0 & 0\\
0 & -1 & 0\\
0 & 0 & -1
\end{bmatrix}$ and $R_{3z} = \begin{bmatrix}
-\frac{1}{2} & -\frac{\sqrt{3}}{2}  & 0\\
\frac{\sqrt{3}}{2} & -\frac{1}{2} & 0\\
0 & 0 & 1
\end{bmatrix}$, respectively.

Under the two-fold rotation, the crystal momenta transform as $ (k_x,k_y,k_z)\rightarrow  (k_x, -k_y, -k_z)$. The Berry curvature and total moment transform as  $(\Omega^{x}_{n(k_x,k_y,k_z)},\Omega^{y}_{n(k_x,k_y,k_z)}, \Omega^{z}_{n(k_x,k_y,k_z)}) \rightarrow (\Omega^{x}_{n(k_x,-k_y,-k_z)},-\Omega^{y}_{n(k_x,-k_y,-k_z)},\\ -\Omega^{z}_{n(k_x,-k_y,-k_z)})$ and $(m^{x}_{n(k_x,k_y,k_z)},m^{y}_{n(k_x,k_y,k_z)}, m^{z}_{n(k_x,k_y,k_z)}) \rightarrow (m^{x}_{n(k_x,-k_y,-k_z)},-m^{y}_{n(k_x,-k_y,-k_z)}, -m^{z}_{n(k_x,-k_y,-k_z)})$. Because this symmetry is present in Tellurium, the Berry curvature and total moment must be invariant under this transformation.

Under the three-fold rotation, the crystal momenta transform as $(k_x,k_y,k_z)\rightarrow \textbf{k}^{'}=(-\frac{1}{2} k_x - \frac{\sqrt{3}}{2} k_y, \frac{\sqrt{3}}{2} k_x - \frac{1}{2} k_y, k_z)$. The Berry curvature transforms as $(\Omega^{x}_{n(k_x,k_y,k_z)},\Omega^{y}_{n(k_x,k_y,k_z)}, \Omega^{z}_{n(k_x,k_y,k_z)})\rightarrow (-\frac{1}{2} \Omega^{x}_{n\textbf{k}^{'}} - \frac{\sqrt{3}}{2}  \Omega^{y}_{n\textbf{k}^{'}}, \frac{\sqrt{3}}{2}  \Omega^{x}_{n\textbf{k}^{'}} - \frac{1}{2}  \Omega^{y}_{n\textbf{k}^{'}}, \Omega^{z}_{n\textbf{k}^{'}})$. The total moment transforms in an identical manner. Symmetry enforces these quantities to remain invariant under this transformation.

For CoSi (SG-198), we select the following set of group generators: $\{C_{2z} | \frac{1}{2} 0 \frac{1}{2} \}$, $\{C_{2y} | 0 \frac{1}{2} \frac{1}{2}\}$, and $\{C_{3_{111}} | 0 0 0\}$. These symmetries impose the following constraints on the Berry curvature of a Bloch electron. \textbf{$C_{2z}$}:  $(\Omega^{x}_{n(k_x,k_y,k_z)},\Omega^{y}_{n(k_x,k_y,k_z)}, \Omega^{z}_{n(k_x,k_y,k_z)}) = (-\Omega^{x}_{n(-k_x,-k_y,k_z)},-\Omega^{y}_{n(-k_x,-k_y,k_z)},\Omega^{z}_{n(-k_x,-k_y,k_z)})$, \textbf{$C_{2y}$}: $(\Omega^{x}_{n(k_x,k_y,k_z)},\Omega^{y}_{n(k_x,k_y,k_z)}, \Omega^{z}_{n(k_x,k_y,k_z)}) = (-\Omega^{x}_{n(-k_x,k_y,-k_z)},\Omega^{y}_{n(-k_x,k_y,-k_z)},-\Omega^{z}_{n(-k_x,k_y,-k_z)})$, and \textbf{$C_{3_{111}}$}: $(\Omega^{x}_{n(k_x,k_y,k_z)},\Omega^{y}_{n(k_x,k_y,k_z)}, \Omega^{z}_{n(k_x,k_y,k_z)}) = (\Omega^{z}_{n(k_y,k_z,k_x)},\Omega^{x}_{n(k_y,k_z,k_x)},\Omega^{y}_{n(k_y,k_z,k_x)})$. Identical constraints exist on the total magnetic moment of a Bloch electron.

\end{widetext}

\bibliographystyle{apsrev}
\bibliography{manuscript.bib} 

\begin{thebibliography}{54}
\expandafter\ifx\csname natexlab\endcsname\relax\def\natexlab#1{#1}\fi
\expandafter\ifx\csname bibnamefont\endcsname\relax
  \def\bibnamefont#1{#1}\fi
\expandafter\ifx\csname bibfnamefont\endcsname\relax
  \def\bibfnamefont#1{#1}\fi
\expandafter\ifx\csname citenamefont\endcsname\relax
  \def\citenamefont#1{#1}\fi
\expandafter\ifx\csname url\endcsname\relax
  \def\url#1{\texttt{#1}}\fi
\expandafter\ifx\csname urlprefix\endcsname\relax\def\urlprefix{URL }\fi
\providecommand{\bibinfo}[2]{#2}
\providecommand{\eprint}[2][]{\url{#2}}

\bibitem[{\citenamefont{Xiao et~al.}(2010)\citenamefont{Xiao, Chang, and Niu}}]{RevModPhys.82.1959}
\bibinfo{author}{\bibfnamefont{D.}~\bibnamefont{Xiao}}, \bibinfo{author}{\bibfnamefont{M.-C.} \bibnamefont{Chang}}, \bibnamefont{and} \bibinfo{author}{\bibfnamefont{Q.}~\bibnamefont{Niu}}, \bibinfo{journal}{Rev. Mod. Phys.} \textbf{\bibinfo{volume}{82}}, \bibinfo{pages}{1959} (\bibinfo{year}{2010}), \urlprefix\url{https://link.aps.org/doi/10.1103/RevModPhys.82.1959}.

\bibitem[{\citenamefont{Ma et~al.}(2023)\citenamefont{Ma, Krishna~Kumar, Xu, Koppens, and Song}}]{Ma2023}
\bibinfo{author}{\bibfnamefont{Q.}~\bibnamefont{Ma}}, \bibinfo{author}{\bibfnamefont{R.}~\bibnamefont{Krishna~Kumar}}, \bibinfo{author}{\bibfnamefont{S.-Y.} \bibnamefont{Xu}}, \bibinfo{author}{\bibfnamefont{F.~H.~L.} \bibnamefont{Koppens}}, \bibnamefont{and} \bibinfo{author}{\bibfnamefont{J.~C.~W.} \bibnamefont{Song}}, \bibinfo{journal}{Nature Reviews Physics} \textbf{\bibinfo{volume}{5}}, \bibinfo{pages}{170–184} (\bibinfo{year}{2023}), ISSN \bibinfo{issn}{2522-5820}, \urlprefix\url{http://dx.doi.org/10.1038/s42254-022-00551-2}.

\bibitem[{\citenamefont{Rappoport et~al.}(2023)\citenamefont{Rappoport, Morgado, Lanneb\`ere, and Silveirinha}}]{PhysRevLett.130.076901}
\bibinfo{author}{\bibfnamefont{T.~G.} \bibnamefont{Rappoport}}, \bibinfo{author}{\bibfnamefont{T.~A.} \bibnamefont{Morgado}}, \bibinfo{author}{\bibfnamefont{S.}~\bibnamefont{Lanneb\`ere}}, \bibnamefont{and} \bibinfo{author}{\bibfnamefont{M.~G.} \bibnamefont{Silveirinha}}, \bibinfo{journal}{Phys. Rev. Lett.} \textbf{\bibinfo{volume}{130}}, \bibinfo{pages}{076901} (\bibinfo{year}{2023}), \urlprefix\url{https://link.aps.org/doi/10.1103/PhysRevLett.130.076901}.

\bibitem[{\citenamefont{K\"onig et~al.}(2019)\citenamefont{K\"onig, Dzero, Levchenko, and Pesin}}]{PhysRevB.99.155404}
\bibinfo{author}{\bibfnamefont{E.~J.} \bibnamefont{K\"onig}}, \bibinfo{author}{\bibfnamefont{M.}~\bibnamefont{Dzero}}, \bibinfo{author}{\bibfnamefont{A.}~\bibnamefont{Levchenko}}, \bibnamefont{and} \bibinfo{author}{\bibfnamefont{D.~A.} \bibnamefont{Pesin}}, \bibinfo{journal}{Phys. Rev. B} \textbf{\bibinfo{volume}{99}}, \bibinfo{pages}{155404} (\bibinfo{year}{2019}), \urlprefix\url{https://link.aps.org/doi/10.1103/PhysRevB.99.155404}.

\bibitem[{\citenamefont{Sodemann and Fu}(2015)}]{PhysRevLett.115.216806}
\bibinfo{author}{\bibfnamefont{I.}~\bibnamefont{Sodemann}} \bibnamefont{and} \bibinfo{author}{\bibfnamefont{L.}~\bibnamefont{Fu}}, \bibinfo{journal}{Phys. Rev. Lett.} \textbf{\bibinfo{volume}{115}}, \bibinfo{pages}{216806} (\bibinfo{year}{2015}), \urlprefix\url{https://link.aps.org/doi/10.1103/PhysRevLett.115.216806}.

\bibitem[{\citenamefont{de~Juan et~al.}(2017)\citenamefont{de~Juan, Grushin, Morimoto, and Moore}}]{deJuan2017}
\bibinfo{author}{\bibfnamefont{F.}~\bibnamefont{de~Juan}}, \bibinfo{author}{\bibfnamefont{A.~G.} \bibnamefont{Grushin}}, \bibinfo{author}{\bibfnamefont{T.}~\bibnamefont{Morimoto}}, \bibnamefont{and} \bibinfo{author}{\bibfnamefont{J.~E.} \bibnamefont{Moore}}, \bibinfo{journal}{Nature Communications} \textbf{\bibinfo{volume}{8}} (\bibinfo{year}{2017}), ISSN \bibinfo{issn}{2041-1723}, \urlprefix\url{http://dx.doi.org/10.1038/ncomms15995}.

\bibitem[{\citenamefont{Tsirkin et~al.}(2018)\citenamefont{Tsirkin, Puente, and Souza}}]{PhysRevB.97.035158}
\bibinfo{author}{\bibfnamefont{S.~S.} \bibnamefont{Tsirkin}}, \bibinfo{author}{\bibfnamefont{P.~A.} \bibnamefont{Puente}}, \bibnamefont{and} \bibinfo{author}{\bibfnamefont{I.}~\bibnamefont{Souza}}, \bibinfo{journal}{Phys. Rev. B} \textbf{\bibinfo{volume}{97}}, \bibinfo{pages}{035158} (\bibinfo{year}{2018}), \urlprefix\url{https://link.aps.org/doi/10.1103/PhysRevB.97.035158}.

\bibitem[{\citenamefont{\ifmmode~\mbox{\c{S}}\else \c{S}\fi{}ahin et~al.}(2018)\citenamefont{\ifmmode~\mbox{\c{S}}\else \c{S}\fi{}ahin, Rou, Ma, and Pesin}}]{PhysRevB.97.205206}
\bibinfo{author}{\bibfnamefont{C.}~\bibnamefont{\ifmmode~\mbox{\c{S}}\else \c{S}\fi{}ahin}}, \bibinfo{author}{\bibfnamefont{J.}~\bibnamefont{Rou}}, \bibinfo{author}{\bibfnamefont{J.}~\bibnamefont{Ma}}, \bibnamefont{and} \bibinfo{author}{\bibfnamefont{D.~A.} \bibnamefont{Pesin}}, \bibinfo{journal}{Phys. Rev. B} \textbf{\bibinfo{volume}{97}}, \bibinfo{pages}{205206} (\bibinfo{year}{2018}), \urlprefix\url{https://link.aps.org/doi/10.1103/PhysRevB.97.205206}.

\bibitem[{\citenamefont{Ma and Pesin}(2015)}]{PhysRevB.92.235205}
\bibinfo{author}{\bibfnamefont{J.}~\bibnamefont{Ma}} \bibnamefont{and} \bibinfo{author}{\bibfnamefont{D.~A.} \bibnamefont{Pesin}}, \bibinfo{journal}{Phys. Rev. B} \textbf{\bibinfo{volume}{92}}, \bibinfo{pages}{235205} (\bibinfo{year}{2015}), \urlprefix\url{https://link.aps.org/doi/10.1103/PhysRevB.92.235205}.

\bibitem[{\citenamefont{Li et~al.}(2022)\citenamefont{Li, Gao, Gu, Zhang, Iitaka, and Liu}}]{PhysRevB.105.125201}
\bibinfo{author}{\bibfnamefont{Z.}~\bibnamefont{Li}}, \bibinfo{author}{\bibfnamefont{Y.}~\bibnamefont{Gao}}, \bibinfo{author}{\bibfnamefont{Y.}~\bibnamefont{Gu}}, \bibinfo{author}{\bibfnamefont{S.}~\bibnamefont{Zhang}}, \bibinfo{author}{\bibfnamefont{T.}~\bibnamefont{Iitaka}}, \bibnamefont{and} \bibinfo{author}{\bibfnamefont{W.~M.} \bibnamefont{Liu}}, \bibinfo{journal}{Phys. Rev. B} \textbf{\bibinfo{volume}{105}}, \bibinfo{pages}{125201} (\bibinfo{year}{2022}), \urlprefix\url{https://link.aps.org/doi/10.1103/PhysRevB.105.125201}.

\bibitem[{\citenamefont{Hakimi et~al.}(2023)\citenamefont{Hakimi, Rouhi, Rappoport, Silveirinha, and Capolino}}]{https://doi.org/10.48550/arxiv.2312.15142}
\bibinfo{author}{\bibfnamefont{A.}~\bibnamefont{Hakimi}}, \bibinfo{author}{\bibfnamefont{K.}~\bibnamefont{Rouhi}}, \bibinfo{author}{\bibfnamefont{T.~G.} \bibnamefont{Rappoport}}, \bibinfo{author}{\bibfnamefont{M.~G.} \bibnamefont{Silveirinha}}, \bibnamefont{and} \bibinfo{author}{\bibfnamefont{F.}~\bibnamefont{Capolino}}, \emph{\bibinfo{title}{Chiral terahertz lasing with berry curvature dipoles}} (\bibinfo{year}{2023}), \urlprefix\url{https://arxiv.org/abs/2312.15142}.

\bibitem[{\citenamefont{Morgado et~al.}(2024)\citenamefont{Morgado, Rappoport, Tsirkin, Lannebère, Souza, and Silveirinha}}]{https://doi.org/10.48550/arxiv.2401.13764}
\bibinfo{author}{\bibfnamefont{T.~A.} \bibnamefont{Morgado}}, \bibinfo{author}{\bibfnamefont{T.~G.} \bibnamefont{Rappoport}}, \bibinfo{author}{\bibfnamefont{S.~S.} \bibnamefont{Tsirkin}}, \bibinfo{author}{\bibfnamefont{S.}~\bibnamefont{Lannebère}}, \bibinfo{author}{\bibfnamefont{I.}~\bibnamefont{Souza}}, \bibnamefont{and} \bibinfo{author}{\bibfnamefont{M.~G.} \bibnamefont{Silveirinha}}, \emph{\bibinfo{title}{Non-hermitian linear electrooptic effect in 3d materials}} (\bibinfo{year}{2024}), \urlprefix\url{https://arxiv.org/abs/2401.13764}.

\bibitem[{\citenamefont{Shi et~al.}(2023)\citenamefont{Shi, Matsyshyn, Song, and Villadiego}}]{PhysRevB.107.125151}
\bibinfo{author}{\bibfnamefont{L.-k.} \bibnamefont{Shi}}, \bibinfo{author}{\bibfnamefont{O.}~\bibnamefont{Matsyshyn}}, \bibinfo{author}{\bibfnamefont{J.~C.~W.} \bibnamefont{Song}}, \bibnamefont{and} \bibinfo{author}{\bibfnamefont{I.~S.} \bibnamefont{Villadiego}}, \bibinfo{journal}{Phys. Rev. B} \textbf{\bibinfo{volume}{107}}, \bibinfo{pages}{125151} (\bibinfo{year}{2023}), \urlprefix\url{https://link.aps.org/doi/10.1103/PhysRevB.107.125151}.

\bibitem[{\citenamefont{Dai and Rappe}(2023)}]{Dai2023}
\bibinfo{author}{\bibfnamefont{Z.}~\bibnamefont{Dai}} \bibnamefont{and} \bibinfo{author}{\bibfnamefont{A.~M.} \bibnamefont{Rappe}}, \bibinfo{journal}{Chemical Physics Reviews} \textbf{\bibinfo{volume}{4}} (\bibinfo{year}{2023}), ISSN \bibinfo{issn}{2688-4070}, \urlprefix\url{http://dx.doi.org/10.1063/5.0101513}.

\bibitem[{\citenamefont{Moore and Orenstein}(2010)}]{PhysRevLett.105.026805}
\bibinfo{author}{\bibfnamefont{J.~E.} \bibnamefont{Moore}} \bibnamefont{and} \bibinfo{author}{\bibfnamefont{J.}~\bibnamefont{Orenstein}}, \bibinfo{journal}{Phys. Rev. Lett.} \textbf{\bibinfo{volume}{105}}, \bibinfo{pages}{026805} (\bibinfo{year}{2010}), \urlprefix\url{https://link.aps.org/doi/10.1103/PhysRevLett.105.026805}.

\bibitem[{\citenamefont{Golub et~al.}(2020)\citenamefont{Golub, Ivchenko, and Spivak}}]{PhysRevB.102.085202}
\bibinfo{author}{\bibfnamefont{L.~E.} \bibnamefont{Golub}}, \bibinfo{author}{\bibfnamefont{E.~L.} \bibnamefont{Ivchenko}}, \bibnamefont{and} \bibinfo{author}{\bibfnamefont{B.}~\bibnamefont{Spivak}}, \bibinfo{journal}{Phys. Rev. B} \textbf{\bibinfo{volume}{102}}, \bibinfo{pages}{085202} (\bibinfo{year}{2020}), \urlprefix\url{https://link.aps.org/doi/10.1103/PhysRevB.102.085202}.

\bibitem[{\citenamefont{Low et~al.}(2015)\citenamefont{Low, Jiang, and Guinea}}]{PhysRevB.92.235447}
\bibinfo{author}{\bibfnamefont{T.}~\bibnamefont{Low}}, \bibinfo{author}{\bibfnamefont{Y.}~\bibnamefont{Jiang}}, \bibnamefont{and} \bibinfo{author}{\bibfnamefont{F.}~\bibnamefont{Guinea}}, \bibinfo{journal}{Phys. Rev. B} \textbf{\bibinfo{volume}{92}}, \bibinfo{pages}{235447} (\bibinfo{year}{2015}), \urlprefix\url{https://link.aps.org/doi/10.1103/PhysRevB.92.235447}.

\bibitem[{\citenamefont{Thonhauser et~al.}(2005)\citenamefont{Thonhauser, Ceresoli, Vanderbilt, and Resta}}]{PhysRevLett.95.137205}
\bibinfo{author}{\bibfnamefont{T.}~\bibnamefont{Thonhauser}}, \bibinfo{author}{\bibfnamefont{D.}~\bibnamefont{Ceresoli}}, \bibinfo{author}{\bibfnamefont{D.}~\bibnamefont{Vanderbilt}}, \bibnamefont{and} \bibinfo{author}{\bibfnamefont{R.}~\bibnamefont{Resta}}, \bibinfo{journal}{Phys. Rev. Lett.} \textbf{\bibinfo{volume}{95}}, \bibinfo{pages}{137205} (\bibinfo{year}{2005}), \urlprefix\url{https://link.aps.org/doi/10.1103/PhysRevLett.95.137205}.

\bibitem[{\citenamefont{Lopez et~al.}(2012)\citenamefont{Lopez, Vanderbilt, Thonhauser, and Souza}}]{PhysRevB.85.014435}
\bibinfo{author}{\bibfnamefont{M.~G.} \bibnamefont{Lopez}}, \bibinfo{author}{\bibfnamefont{D.}~\bibnamefont{Vanderbilt}}, \bibinfo{author}{\bibfnamefont{T.}~\bibnamefont{Thonhauser}}, \bibnamefont{and} \bibinfo{author}{\bibfnamefont{I.}~\bibnamefont{Souza}}, \bibinfo{journal}{Phys. Rev. B} \textbf{\bibinfo{volume}{85}}, \bibinfo{pages}{014435} (\bibinfo{year}{2012}), \urlprefix\url{https://link.aps.org/doi/10.1103/PhysRevB.85.014435}.

\bibitem[{\citenamefont{Morimoto et~al.}(2016)\citenamefont{Morimoto, Zhong, Orenstein, and Moore}}]{PhysRevB.94.245121}
\bibinfo{author}{\bibfnamefont{T.}~\bibnamefont{Morimoto}}, \bibinfo{author}{\bibfnamefont{S.}~\bibnamefont{Zhong}}, \bibinfo{author}{\bibfnamefont{J.}~\bibnamefont{Orenstein}}, \bibnamefont{and} \bibinfo{author}{\bibfnamefont{J.~E.} \bibnamefont{Moore}}, \bibinfo{journal}{Phys. Rev. B} \textbf{\bibinfo{volume}{94}}, \bibinfo{pages}{245121} (\bibinfo{year}{2016}), \urlprefix\url{https://link.aps.org/doi/10.1103/PhysRevB.94.245121}.

\bibitem[{\citenamefont{Malashevich et~al.}(2010)\citenamefont{Malashevich, Souza, Coh, and Vanderbilt}}]{Malashevich2010}
\bibinfo{author}{\bibfnamefont{A.}~\bibnamefont{Malashevich}}, \bibinfo{author}{\bibfnamefont{I.}~\bibnamefont{Souza}}, \bibinfo{author}{\bibfnamefont{S.}~\bibnamefont{Coh}}, \bibnamefont{and} \bibinfo{author}{\bibfnamefont{D.}~\bibnamefont{Vanderbilt}}, \bibinfo{journal}{New Journal of Physics} \textbf{\bibinfo{volume}{12}}, \bibinfo{pages}{053032} (\bibinfo{year}{2010}), ISSN \bibinfo{issn}{1367-2630}, \urlprefix\url{http://dx.doi.org/10.1088/1367-2630/12/5/053032}.

\bibitem[{\citenamefont{Vanderbilt}(2018)}]{Vanderbilt2018}
\bibinfo{author}{\bibfnamefont{D.}~\bibnamefont{Vanderbilt}}, \emph{\bibinfo{title}{Berry Phases in Electronic Structure Theory: Electric Polarization, Orbital Magnetization and Topological Insulators}} (\bibinfo{publisher}{Cambridge University Press}, \bibinfo{year}{2018}), ISBN \bibinfo{isbn}{9781107157651}, \urlprefix\url{http://dx.doi.org/10.1017/9781316662205}.

\bibitem[{\citenamefont{Rou et~al.}(2017)\citenamefont{Rou, \ifmmode~\mbox{\c{S}}\else \c{S}\fi{}ahin, Ma, and Pesin}}]{PhysRevB.96.035120}
\bibinfo{author}{\bibfnamefont{J.}~\bibnamefont{Rou}}, \bibinfo{author}{\bibfnamefont{C.}~\bibnamefont{\ifmmode~\mbox{\c{S}}\else \c{S}\fi{}ahin}}, \bibinfo{author}{\bibfnamefont{J.}~\bibnamefont{Ma}}, \bibnamefont{and} \bibinfo{author}{\bibfnamefont{D.~A.} \bibnamefont{Pesin}}, \bibinfo{journal}{Phys. Rev. B} \textbf{\bibinfo{volume}{96}}, \bibinfo{pages}{035120} (\bibinfo{year}{2017}), \urlprefix\url{https://link.aps.org/doi/10.1103/PhysRevB.96.035120}.

\bibitem[{\citenamefont{Zhong et~al.}(2016)\citenamefont{Zhong, Moore, and Souza}}]{PhysRevLett.116.077201}
\bibinfo{author}{\bibfnamefont{S.}~\bibnamefont{Zhong}}, \bibinfo{author}{\bibfnamefont{J.~E.} \bibnamefont{Moore}}, \bibnamefont{and} \bibinfo{author}{\bibfnamefont{I.}~\bibnamefont{Souza}}, \bibinfo{journal}{Phys. Rev. Lett.} \textbf{\bibinfo{volume}{116}}, \bibinfo{pages}{077201} (\bibinfo{year}{2016}), \urlprefix\url{https://link.aps.org/doi/10.1103/PhysRevLett.116.077201}.

\bibitem[{\citenamefont{Wang et~al.}(2020)\citenamefont{Wang, Morimoto, and Moore}}]{PhysRevB.101.174419}
\bibinfo{author}{\bibfnamefont{Y.-Q.} \bibnamefont{Wang}}, \bibinfo{author}{\bibfnamefont{T.}~\bibnamefont{Morimoto}}, \bibnamefont{and} \bibinfo{author}{\bibfnamefont{J.~E.} \bibnamefont{Moore}}, \bibinfo{journal}{Phys. Rev. B} \textbf{\bibinfo{volume}{101}}, \bibinfo{pages}{174419} (\bibinfo{year}{2020}), \urlprefix\url{https://link.aps.org/doi/10.1103/PhysRevB.101.174419}.

\bibitem[{\citenamefont{Hornreich and Shtrikman}(1968)}]{PhysRev.171.1065}
\bibinfo{author}{\bibfnamefont{R.~M.} \bibnamefont{Hornreich}} \bibnamefont{and} \bibinfo{author}{\bibfnamefont{S.}~\bibnamefont{Shtrikman}}, \bibinfo{journal}{Phys. Rev.} \textbf{\bibinfo{volume}{171}}, \bibinfo{pages}{1065} (\bibinfo{year}{1968}), \urlprefix\url{https://link.aps.org/doi/10.1103/PhysRev.171.1065}.

\bibitem[{\citenamefont{Malashevich and Souza}(2010)}]{PhysRevB.82.245118}
\bibinfo{author}{\bibfnamefont{A.}~\bibnamefont{Malashevich}} \bibnamefont{and} \bibinfo{author}{\bibfnamefont{I.}~\bibnamefont{Souza}}, \bibinfo{journal}{Phys. Rev. B} \textbf{\bibinfo{volume}{82}}, \bibinfo{pages}{245118} (\bibinfo{year}{2010}), \urlprefix\url{https://link.aps.org/doi/10.1103/PhysRevB.82.245118}.

\bibitem[{\citenamefont{Flicker et~al.}(2018)\citenamefont{Flicker, de~Juan, Bradlyn, Morimoto, Vergniory, and Grushin}}]{PhysRevB.98.155145}
\bibinfo{author}{\bibfnamefont{F.}~\bibnamefont{Flicker}}, \bibinfo{author}{\bibfnamefont{F.}~\bibnamefont{de~Juan}}, \bibinfo{author}{\bibfnamefont{B.}~\bibnamefont{Bradlyn}}, \bibinfo{author}{\bibfnamefont{T.}~\bibnamefont{Morimoto}}, \bibinfo{author}{\bibfnamefont{M.~G.} \bibnamefont{Vergniory}}, \bibnamefont{and} \bibinfo{author}{\bibfnamefont{A.~G.} \bibnamefont{Grushin}}, \bibinfo{journal}{Phys. Rev. B} \textbf{\bibinfo{volume}{98}}, \bibinfo{pages}{155145} (\bibinfo{year}{2018}), \urlprefix\url{https://link.aps.org/doi/10.1103/PhysRevB.98.155145}.

\bibitem[{198(1987)}]{1987}
\emph{\bibinfo{title}{Electro-optic and Photorefractive Materials: Proceedings of the International School on Material Science and Technology, Erice, Italy, July 6–17, 1986}} (\bibinfo{publisher}{Springer Berlin Heidelberg}, \bibinfo{year}{1987}), ISBN \bibinfo{isbn}{9783642719073}, \urlprefix\url{http://dx.doi.org/10.1007/978-3-642-71907-3}.

\bibitem[{\citenamefont{Weber}(2018)}]{Weber2018}
\bibinfo{author}{\bibfnamefont{M.~J.} \bibnamefont{Weber}}, \emph{\bibinfo{title}{Handbook of Optical Materials}} (\bibinfo{publisher}{CRC Press}, \bibinfo{year}{2018}), ISBN \bibinfo{isbn}{9781315219615}, \urlprefix\url{http://dx.doi.org/10.1201/9781315219615}.

\bibitem[{\citenamefont{Jiang et~al.}(2020)\citenamefont{Jiang, Paillard, Xiang, and Bellaiche}}]{PhysRevLett.125.017401}
\bibinfo{author}{\bibfnamefont{Z.}~\bibnamefont{Jiang}}, \bibinfo{author}{\bibfnamefont{C.}~\bibnamefont{Paillard}}, \bibinfo{author}{\bibfnamefont{H.}~\bibnamefont{Xiang}}, \bibnamefont{and} \bibinfo{author}{\bibfnamefont{L.}~\bibnamefont{Bellaiche}}, \bibinfo{journal}{Phys. Rev. Lett.} \textbf{\bibinfo{volume}{125}}, \bibinfo{pages}{017401} (\bibinfo{year}{2020}), \urlprefix\url{https://link.aps.org/doi/10.1103/PhysRevLett.125.017401}.

\bibitem[{\citenamefont{Qiu et~al.}(2022)\citenamefont{Qiu, Charnas, Niu, Wang, Wu, and Ye}}]{Qiu2022}
\bibinfo{author}{\bibfnamefont{G.}~\bibnamefont{Qiu}}, \bibinfo{author}{\bibfnamefont{A.}~\bibnamefont{Charnas}}, \bibinfo{author}{\bibfnamefont{C.}~\bibnamefont{Niu}}, \bibinfo{author}{\bibfnamefont{Y.}~\bibnamefont{Wang}}, \bibinfo{author}{\bibfnamefont{W.}~\bibnamefont{Wu}}, \bibnamefont{and} \bibinfo{author}{\bibfnamefont{P.~D.} \bibnamefont{Ye}}, \bibinfo{journal}{npj 2D Materials and Applications} \textbf{\bibinfo{volume}{6}} (\bibinfo{year}{2022}), ISSN \bibinfo{issn}{2397-7132}, \urlprefix\url{http://dx.doi.org/10.1038/s41699-022-00293-w}.

\bibitem[{\citenamefont{Sakano et~al.}(2020)\citenamefont{Sakano, Hirayama, Takahashi, Akebi, Nakayama, Kuroda, Taguchi, Yoshikawa, Miyamoto, Okuda et~al.}}]{PhysRevLett.124.136404}
\bibinfo{author}{\bibfnamefont{M.}~\bibnamefont{Sakano}}, \bibinfo{author}{\bibfnamefont{M.}~\bibnamefont{Hirayama}}, \bibinfo{author}{\bibfnamefont{T.}~\bibnamefont{Takahashi}}, \bibinfo{author}{\bibfnamefont{S.}~\bibnamefont{Akebi}}, \bibinfo{author}{\bibfnamefont{M.}~\bibnamefont{Nakayama}}, \bibinfo{author}{\bibfnamefont{K.}~\bibnamefont{Kuroda}}, \bibinfo{author}{\bibfnamefont{K.}~\bibnamefont{Taguchi}}, \bibinfo{author}{\bibfnamefont{T.}~\bibnamefont{Yoshikawa}}, \bibinfo{author}{\bibfnamefont{K.}~\bibnamefont{Miyamoto}}, \bibinfo{author}{\bibfnamefont{T.}~\bibnamefont{Okuda}}, \bibnamefont{et~al.}, \bibinfo{journal}{Phys. Rev. Lett.} \textbf{\bibinfo{volume}{124}}, \bibinfo{pages}{136404} (\bibinfo{year}{2020}), \urlprefix\url{https://link.aps.org/doi/10.1103/PhysRevLett.124.136404}.

\bibitem[{\citenamefont{Gatti et~al.}(2020)\citenamefont{Gatti, Gos\'albez-Mart\'{\i}nez, Tsirkin, Fanciulli, Puppin, Polishchuk, Moser, Testa, Martino, Roth et~al.}}]{PhysRevLett.125.216402}
\bibinfo{author}{\bibfnamefont{G.}~\bibnamefont{Gatti}}, \bibinfo{author}{\bibfnamefont{D.}~\bibnamefont{Gos\'albez-Mart\'{\i}nez}}, \bibinfo{author}{\bibfnamefont{S.~S.} \bibnamefont{Tsirkin}}, \bibinfo{author}{\bibfnamefont{M.}~\bibnamefont{Fanciulli}}, \bibinfo{author}{\bibfnamefont{M.}~\bibnamefont{Puppin}}, \bibinfo{author}{\bibfnamefont{S.}~\bibnamefont{Polishchuk}}, \bibinfo{author}{\bibfnamefont{S.}~\bibnamefont{Moser}}, \bibinfo{author}{\bibfnamefont{L.}~\bibnamefont{Testa}}, \bibinfo{author}{\bibfnamefont{E.}~\bibnamefont{Martino}}, \bibinfo{author}{\bibfnamefont{S.}~\bibnamefont{Roth}}, \bibnamefont{et~al.}, \bibinfo{journal}{Phys. Rev. Lett.} \textbf{\bibinfo{volume}{125}}, \bibinfo{pages}{216402} (\bibinfo{year}{2020}), \urlprefix\url{https://link.aps.org/doi/10.1103/PhysRevLett.125.216402}.

\bibitem[{\citenamefont{Chang et~al.}(2018)\citenamefont{Chang, Wieder, Schindler, Sanchez, Belopolski, Huang, Singh, Wu, Chang, Neupert et~al.}}]{Chang2018}
\bibinfo{author}{\bibfnamefont{G.}~\bibnamefont{Chang}}, \bibinfo{author}{\bibfnamefont{B.~J.} \bibnamefont{Wieder}}, \bibinfo{author}{\bibfnamefont{F.}~\bibnamefont{Schindler}}, \bibinfo{author}{\bibfnamefont{D.~S.} \bibnamefont{Sanchez}}, \bibinfo{author}{\bibfnamefont{I.}~\bibnamefont{Belopolski}}, \bibinfo{author}{\bibfnamefont{S.-M.} \bibnamefont{Huang}}, \bibinfo{author}{\bibfnamefont{B.}~\bibnamefont{Singh}}, \bibinfo{author}{\bibfnamefont{D.}~\bibnamefont{Wu}}, \bibinfo{author}{\bibfnamefont{T.-R.} \bibnamefont{Chang}}, \bibinfo{author}{\bibfnamefont{T.}~\bibnamefont{Neupert}}, \bibnamefont{et~al.}, \bibinfo{journal}{Nature Materials} \textbf{\bibinfo{volume}{17}}, \bibinfo{pages}{978–985} (\bibinfo{year}{2018}), ISSN \bibinfo{issn}{1476-4660}, \urlprefix\url{http://dx.doi.org/10.1038/s41563-018-0169-3}.

\bibitem[{\citenamefont{Bradlyn et~al.}(2016)\citenamefont{Bradlyn, Cano, Wang, Vergniory, Felser, Cava, and Bernevig}}]{Bradlyn2016}
\bibinfo{author}{\bibfnamefont{B.}~\bibnamefont{Bradlyn}}, \bibinfo{author}{\bibfnamefont{J.}~\bibnamefont{Cano}}, \bibinfo{author}{\bibfnamefont{Z.}~\bibnamefont{Wang}}, \bibinfo{author}{\bibfnamefont{M.~G.} \bibnamefont{Vergniory}}, \bibinfo{author}{\bibfnamefont{C.}~\bibnamefont{Felser}}, \bibinfo{author}{\bibfnamefont{R.~J.} \bibnamefont{Cava}}, \bibnamefont{and} \bibinfo{author}{\bibfnamefont{B.~A.} \bibnamefont{Bernevig}}, \bibinfo{journal}{Science} \textbf{\bibinfo{volume}{353}} (\bibinfo{year}{2016}), ISSN \bibinfo{issn}{1095-9203}, \urlprefix\url{http://dx.doi.org/10.1126/science.aaf5037}.

\bibitem[{\citenamefont{Cochran et~al.}(2023)\citenamefont{Cochran, Belopolski, Manna, Yahyavi, Liu, Sanchez, Cheng, Yang, Multer, Yin et~al.}}]{PhysRevLett.130.066402}
\bibinfo{author}{\bibfnamefont{T.~A.} \bibnamefont{Cochran}}, \bibinfo{author}{\bibfnamefont{I.}~\bibnamefont{Belopolski}}, \bibinfo{author}{\bibfnamefont{K.}~\bibnamefont{Manna}}, \bibinfo{author}{\bibfnamefont{M.}~\bibnamefont{Yahyavi}}, \bibinfo{author}{\bibfnamefont{Y.}~\bibnamefont{Liu}}, \bibinfo{author}{\bibfnamefont{D.~S.} \bibnamefont{Sanchez}}, \bibinfo{author}{\bibfnamefont{Z.-J.} \bibnamefont{Cheng}}, \bibinfo{author}{\bibfnamefont{X.~P.} \bibnamefont{Yang}}, \bibinfo{author}{\bibfnamefont{D.}~\bibnamefont{Multer}}, \bibinfo{author}{\bibfnamefont{J.-X.} \bibnamefont{Yin}}, \bibnamefont{et~al.}, \bibinfo{journal}{Phys. Rev. Lett.} \textbf{\bibinfo{volume}{130}}, \bibinfo{pages}{066402} (\bibinfo{year}{2023}), \urlprefix\url{https://link.aps.org/doi/10.1103/PhysRevLett.130.066402}.

\bibitem[{\citenamefont{Rao et~al.}(2019)\citenamefont{Rao, Li, Zhang, Tian, Li, Fu, Tang, Wang, Li, Fan et~al.}}]{Rao2019}
\bibinfo{author}{\bibfnamefont{Z.}~\bibnamefont{Rao}}, \bibinfo{author}{\bibfnamefont{H.}~\bibnamefont{Li}}, \bibinfo{author}{\bibfnamefont{T.}~\bibnamefont{Zhang}}, \bibinfo{author}{\bibfnamefont{S.}~\bibnamefont{Tian}}, \bibinfo{author}{\bibfnamefont{C.}~\bibnamefont{Li}}, \bibinfo{author}{\bibfnamefont{B.}~\bibnamefont{Fu}}, \bibinfo{author}{\bibfnamefont{C.}~\bibnamefont{Tang}}, \bibinfo{author}{\bibfnamefont{L.}~\bibnamefont{Wang}}, \bibinfo{author}{\bibfnamefont{Z.}~\bibnamefont{Li}}, \bibinfo{author}{\bibfnamefont{W.}~\bibnamefont{Fan}}, \bibnamefont{et~al.}, \bibinfo{journal}{Nature} \textbf{\bibinfo{volume}{567}}, \bibinfo{pages}{496–499} (\bibinfo{year}{2019}), ISSN \bibinfo{issn}{1476-4687}, \urlprefix\url{http://dx.doi.org/10.1038/s41586-019-1031-8}.

\bibitem[{\citenamefont{Schr\"{o}ter et~al.}(2019)\citenamefont{Schr\"{o}ter, Pei, Vergniory, Sun, Manna, de~Juan, Krieger, S\"{u}ss, Schmidt, Dudin et~al.}}]{Schrter2019}
\bibinfo{author}{\bibfnamefont{N.~B.~M.} \bibnamefont{Schr\"{o}ter}}, \bibinfo{author}{\bibfnamefont{D.}~\bibnamefont{Pei}}, \bibinfo{author}{\bibfnamefont{M.~G.} \bibnamefont{Vergniory}}, \bibinfo{author}{\bibfnamefont{Y.}~\bibnamefont{Sun}}, \bibinfo{author}{\bibfnamefont{K.}~\bibnamefont{Manna}}, \bibinfo{author}{\bibfnamefont{F.}~\bibnamefont{de~Juan}}, \bibinfo{author}{\bibfnamefont{J.~A.} \bibnamefont{Krieger}}, \bibinfo{author}{\bibfnamefont{V.}~\bibnamefont{S\"{u}ss}}, \bibinfo{author}{\bibfnamefont{M.}~\bibnamefont{Schmidt}}, \bibinfo{author}{\bibfnamefont{P.}~\bibnamefont{Dudin}}, \bibnamefont{et~al.}, \bibinfo{journal}{Nature Physics} \textbf{\bibinfo{volume}{15}}, \bibinfo{pages}{759–765} (\bibinfo{year}{2019}), ISSN \bibinfo{issn}{1745-2481}, \urlprefix\url{http://dx.doi.org/10.1038/s41567-019-0511-y}.

\bibitem[{\citenamefont{Chang et~al.}(2017)\citenamefont{Chang, Xu, Wieder, Sanchez, Huang, Belopolski, Chang, Zhang, Bansil, Lin et~al.}}]{PhysRevLett.119.206401}
\bibinfo{author}{\bibfnamefont{G.}~\bibnamefont{Chang}}, \bibinfo{author}{\bibfnamefont{S.-Y.} \bibnamefont{Xu}}, \bibinfo{author}{\bibfnamefont{B.~J.} \bibnamefont{Wieder}}, \bibinfo{author}{\bibfnamefont{D.~S.} \bibnamefont{Sanchez}}, \bibinfo{author}{\bibfnamefont{S.-M.} \bibnamefont{Huang}}, \bibinfo{author}{\bibfnamefont{I.}~\bibnamefont{Belopolski}}, \bibinfo{author}{\bibfnamefont{T.-R.} \bibnamefont{Chang}}, \bibinfo{author}{\bibfnamefont{S.}~\bibnamefont{Zhang}}, \bibinfo{author}{\bibfnamefont{A.}~\bibnamefont{Bansil}}, \bibinfo{author}{\bibfnamefont{H.}~\bibnamefont{Lin}}, \bibnamefont{et~al.}, \bibinfo{journal}{Phys. Rev. Lett.} \textbf{\bibinfo{volume}{119}}, \bibinfo{pages}{206401} (\bibinfo{year}{2017}), \urlprefix\url{https://link.aps.org/doi/10.1103/PhysRevLett.119.206401}.

\bibitem[{\citenamefont{Ascencio et~al.}(2023)\citenamefont{Ascencio, Jiang, de~Sousa, Lee, Wang, and Low}}]{PhysRevB.108.L201404}
\bibinfo{author}{\bibfnamefont{C.~O.} \bibnamefont{Ascencio}}, \bibinfo{author}{\bibfnamefont{W.}~\bibnamefont{Jiang}}, \bibinfo{author}{\bibfnamefont{D.~J.~P.} \bibnamefont{de~Sousa}}, \bibinfo{author}{\bibfnamefont{S.}~\bibnamefont{Lee}}, \bibinfo{author}{\bibfnamefont{J.-P.} \bibnamefont{Wang}}, \bibnamefont{and} \bibinfo{author}{\bibfnamefont{T.}~\bibnamefont{Low}}, \bibinfo{journal}{Phys. Rev. B} \textbf{\bibinfo{volume}{108}}, \bibinfo{pages}{L201404} (\bibinfo{year}{2023}), \urlprefix\url{https://link.aps.org/doi/10.1103/PhysRevB.108.L201404}.

\bibitem[{\citenamefont{Pshenay-Severin et~al.}(2018)\citenamefont{Pshenay-Severin, Ivanov, Burkov, and Burkov}}]{PshenaySeverin2018}
\bibinfo{author}{\bibfnamefont{D.~A.} \bibnamefont{Pshenay-Severin}}, \bibinfo{author}{\bibfnamefont{Y.~V.} \bibnamefont{Ivanov}}, \bibinfo{author}{\bibfnamefont{A.~A.} \bibnamefont{Burkov}}, \bibnamefont{and} \bibinfo{author}{\bibfnamefont{A.~T.} \bibnamefont{Burkov}}, \bibinfo{journal}{Journal of Physics: Condensed Matter} \textbf{\bibinfo{volume}{30}}, \bibinfo{pages}{135501} (\bibinfo{year}{2018}), ISSN \bibinfo{issn}{1361-648X}, \urlprefix\url{http://dx.doi.org/10.1088/1361-648X/aab0ba}.

\bibitem[{\citenamefont{Tang et~al.}(2017)\citenamefont{Tang, Zhou, and Zhang}}]{PhysRevLett.119.206402}
\bibinfo{author}{\bibfnamefont{P.}~\bibnamefont{Tang}}, \bibinfo{author}{\bibfnamefont{Q.}~\bibnamefont{Zhou}}, \bibnamefont{and} \bibinfo{author}{\bibfnamefont{S.-C.} \bibnamefont{Zhang}}, \bibinfo{journal}{Phys. Rev. Lett.} \textbf{\bibinfo{volume}{119}}, \bibinfo{pages}{206402} (\bibinfo{year}{2017}), \urlprefix\url{https://link.aps.org/doi/10.1103/PhysRevLett.119.206402}.

\bibitem[{\citenamefont{Son and Spivak}(2013)}]{PhysRevB.88.104412}
\bibinfo{author}{\bibfnamefont{D.~T.} \bibnamefont{Son}} \bibnamefont{and} \bibinfo{author}{\bibfnamefont{B.~Z.} \bibnamefont{Spivak}}, \bibinfo{journal}{Phys. Rev. B} \textbf{\bibinfo{volume}{88}}, \bibinfo{pages}{104412} (\bibinfo{year}{2013}), \urlprefix\url{https://link.aps.org/doi/10.1103/PhysRevB.88.104412}.

\bibitem[{\citenamefont{Yang et~al.}(2023)\citenamefont{Yang, Xiao, Robredo, Vergniory, Yan, and Felser}}]{Yang2023}
\bibinfo{author}{\bibfnamefont{Q.}~\bibnamefont{Yang}}, \bibinfo{author}{\bibfnamefont{J.}~\bibnamefont{Xiao}}, \bibinfo{author}{\bibfnamefont{I.}~\bibnamefont{Robredo}}, \bibinfo{author}{\bibfnamefont{M.~G.} \bibnamefont{Vergniory}}, \bibinfo{author}{\bibfnamefont{B.}~\bibnamefont{Yan}}, \bibnamefont{and} \bibinfo{author}{\bibfnamefont{C.}~\bibnamefont{Felser}}, \bibinfo{journal}{Proceedings of the National Academy of Sciences} \textbf{\bibinfo{volume}{120}} (\bibinfo{year}{2023}), ISSN \bibinfo{issn}{1091-6490}, \urlprefix\url{http://dx.doi.org/10.1073/pnas.2305541120}.

\bibitem[{\citenamefont{Go et~al.}(2017)\citenamefont{Go, Hanke, Buhl, Freimuth, Bihlmayer, Lee, Mokrousov, and Bl\"{u}gel}}]{Go2017}
\bibinfo{author}{\bibfnamefont{D.}~\bibnamefont{Go}}, \bibinfo{author}{\bibfnamefont{J.-P.} \bibnamefont{Hanke}}, \bibinfo{author}{\bibfnamefont{P.~M.} \bibnamefont{Buhl}}, \bibinfo{author}{\bibfnamefont{F.}~\bibnamefont{Freimuth}}, \bibinfo{author}{\bibfnamefont{G.}~\bibnamefont{Bihlmayer}}, \bibinfo{author}{\bibfnamefont{H.-W.} \bibnamefont{Lee}}, \bibinfo{author}{\bibfnamefont{Y.}~\bibnamefont{Mokrousov}}, \bibnamefont{and} \bibinfo{author}{\bibfnamefont{S.}~\bibnamefont{Bl\"{u}gel}}, \bibinfo{journal}{Scientific Reports} \textbf{\bibinfo{volume}{7}} (\bibinfo{year}{2017}), ISSN \bibinfo{issn}{2045-2322}, \urlprefix\url{http://dx.doi.org/10.1038/srep46742}.

\bibitem[{\citenamefont{Kim et~al.}(2014)\citenamefont{Kim, Kim, and Sasaki}}]{PhysRevB.89.195137}
\bibinfo{author}{\bibfnamefont{K.-S.} \bibnamefont{Kim}}, \bibinfo{author}{\bibfnamefont{H.-J.} \bibnamefont{Kim}}, \bibnamefont{and} \bibinfo{author}{\bibfnamefont{M.}~\bibnamefont{Sasaki}}, \bibinfo{journal}{Phys. Rev. B} \textbf{\bibinfo{volume}{89}}, \bibinfo{pages}{195137} (\bibinfo{year}{2014}), \urlprefix\url{https://link.aps.org/doi/10.1103/PhysRevB.89.195137}.

\bibitem[{\citenamefont{Amitani and Nishida}(2023)}]{PhysRevB.107.014302}
\bibinfo{author}{\bibfnamefont{T.}~\bibnamefont{Amitani}} \bibnamefont{and} \bibinfo{author}{\bibfnamefont{Y.}~\bibnamefont{Nishida}}, \bibinfo{journal}{Phys. Rev. B} \textbf{\bibinfo{volume}{107}}, \bibinfo{pages}{014302} (\bibinfo{year}{2023}), \urlprefix\url{https://link.aps.org/doi/10.1103/PhysRevB.107.014302}.

\bibitem[{\citenamefont{Fukushima et~al.}(2008)\citenamefont{Fukushima, Kharzeev, and Warringa}}]{PhysRevD.78.074033}
\bibinfo{author}{\bibfnamefont{K.}~\bibnamefont{Fukushima}}, \bibinfo{author}{\bibfnamefont{D.~E.} \bibnamefont{Kharzeev}}, \bibnamefont{and} \bibinfo{author}{\bibfnamefont{H.~J.} \bibnamefont{Warringa}}, \bibinfo{journal}{Phys. Rev. D} \textbf{\bibinfo{volume}{78}}, \bibinfo{pages}{074033} (\bibinfo{year}{2008}), \urlprefix\url{https://link.aps.org/doi/10.1103/PhysRevD.78.074033}.

\bibitem[{\citenamefont{Vazifeh and Franz}(2013)}]{PhysRevLett.111.027201}
\bibinfo{author}{\bibfnamefont{M.~M.} \bibnamefont{Vazifeh}} \bibnamefont{and} \bibinfo{author}{\bibfnamefont{M.}~\bibnamefont{Franz}}, \bibinfo{journal}{Phys. Rev. Lett.} \textbf{\bibinfo{volume}{111}}, \bibinfo{pages}{027201} (\bibinfo{year}{2013}), \urlprefix\url{https://link.aps.org/doi/10.1103/PhysRevLett.111.027201}.

\bibitem[{\citenamefont{Goswami et~al.}(2015)\citenamefont{Goswami, Sharma, and Tewari}}]{PhysRevB.92.161110}
\bibinfo{author}{\bibfnamefont{P.}~\bibnamefont{Goswami}}, \bibinfo{author}{\bibfnamefont{G.}~\bibnamefont{Sharma}}, \bibnamefont{and} \bibinfo{author}{\bibfnamefont{S.}~\bibnamefont{Tewari}}, \bibinfo{journal}{Phys. Rev. B} \textbf{\bibinfo{volume}{92}}, \bibinfo{pages}{161110} (\bibinfo{year}{2015}), \urlprefix\url{https://link.aps.org/doi/10.1103/PhysRevB.92.161110}.

\bibitem[{\citenamefont{Calavalle et~al.}(2022)\citenamefont{Calavalle, Suárez-Rodríguez, Martín-García, Johansson, Vaz, Yang, Maznichenko, Ostanin, Mateo-Alonso, Chuvilin et~al.}}]{Calavalle2022}
\bibinfo{author}{\bibfnamefont{F.}~\bibnamefont{Calavalle}}, \bibinfo{author}{\bibfnamefont{M.}~\bibnamefont{Suárez-Rodríguez}}, \bibinfo{author}{\bibfnamefont{B.}~\bibnamefont{Martín-García}}, \bibinfo{author}{\bibfnamefont{A.}~\bibnamefont{Johansson}}, \bibinfo{author}{\bibfnamefont{D.~C.} \bibnamefont{Vaz}}, \bibinfo{author}{\bibfnamefont{H.}~\bibnamefont{Yang}}, \bibinfo{author}{\bibfnamefont{I.~V.} \bibnamefont{Maznichenko}}, \bibinfo{author}{\bibfnamefont{S.}~\bibnamefont{Ostanin}}, \bibinfo{author}{\bibfnamefont{A.}~\bibnamefont{Mateo-Alonso}}, \bibinfo{author}{\bibfnamefont{A.}~\bibnamefont{Chuvilin}}, \bibnamefont{et~al.}, \bibinfo{journal}{Nature Materials} \textbf{\bibinfo{volume}{21}}, \bibinfo{pages}{526–532} (\bibinfo{year}{2022}), ISSN \bibinfo{issn}{1476-4660}, \urlprefix\url{http://dx.doi.org/10.1038/s41563-022-01211-7}.

\bibitem[{\citenamefont{Furukawa et~al.}(2017)\citenamefont{Furukawa, Shimokawa, Kobayashi, and Itou}}]{Furukawa2017}
\bibinfo{author}{\bibfnamefont{T.}~\bibnamefont{Furukawa}}, \bibinfo{author}{\bibfnamefont{Y.}~\bibnamefont{Shimokawa}}, \bibinfo{author}{\bibfnamefont{K.}~\bibnamefont{Kobayashi}}, \bibnamefont{and} \bibinfo{author}{\bibfnamefont{T.}~\bibnamefont{Itou}}, \bibinfo{journal}{Nature Communications} \textbf{\bibinfo{volume}{8}}, \bibinfo{pages}{954} (\bibinfo{year}{2017}), ISSN \bibinfo{issn}{2041-1723}, \urlprefix\url{https://doi.org/10.1038/s41467-017-01093-3}.

\bibitem[{\citenamefont{Rikken and Avarvari}(2019)}]{PhysRevB.99.245153}
\bibinfo{author}{\bibfnamefont{G.~L. J.~A.} \bibnamefont{Rikken}} \bibnamefont{and} \bibinfo{author}{\bibfnamefont{N.}~\bibnamefont{Avarvari}}, \bibinfo{journal}{Phys. Rev. B} \textbf{\bibinfo{volume}{99}}, \bibinfo{pages}{245153} (\bibinfo{year}{2019}), \urlprefix\url{https://link.aps.org/doi/10.1103/PhysRevB.99.245153}.

\end{thebibliography}

\end{document}